\documentclass[floatfix,%
 aip,
 amsmath,amssymb,
 reprint,%
]{revtex4-1}

\usepackage{graphicx}
\usepackage{dcolumn}
\usepackage{bm}
\usepackage{xcolor}
\usepackage[dvipsnames]{xcolor}

\usepackage[utf8]{inputenc}
\usepackage[T1]{fontenc}
\usepackage{mathptmx}
\usepackage{etoolbox}
\usepackage{subcaption}

\usepackage{newunicodechar}
\newunicodechar{α}{\alpha}

\newcommand{\lasnex}{Lasnex}
\newcommand{\myvec}[1]{\vec{#1}}

\def\Ti{$\textit{T}_\text{ion}$ }
\def\Te{$\textit{T}_\text{e}$ }
\def\bq\begin{equation}
\def\eq\end{equation}

\newcommand{\redub}{}
\def\redub#1{%
  \@ifnextchar_%
    {\@redub{#1}}
    {\@latex@warning{Missing argument for \string\redub}\@redub{#1}_{}}%
}
\def\@redub#1_#2{%
    \colorlet{currentcolor}{.}%
    \color{red}%
    \underbrace{\color{currentcolor}#1}_{\color{red}#2}%
    \color{currentcolor}%
}

\makeatletter
\def\@email#1#2{%
 \endgroup
 \patchcmd{\titleblock@produce}
  {\frontmatter@RRAPformat}
  {\frontmatter@RRAPformat{\produce@RRAP{*#1\href{mailto:#2}{#2}}}\frontmatter@RRAPformat}
  {}{}
}%
\makeatother
\begin{document}

\preprint{AIP/123-QED}

\title[]{Integrated Radiation-Magneto-Hydrodynamic Simulations of Magnetized Burning Plasmas. I. Magnetizing Ignition-Class Designs}

\author{B. Z. Djordjevi\'{c}}
\affiliation{ Lawrence Livermore National Laboratory, Livermore, California 94550, USA}%
 \email{djordjevic3@llnl.gov}
\author{D. J. Strozzi}%
\affiliation{ Lawrence Livermore National Laboratory, Livermore, California 94550, USA}%
\author{G. B. Zimmerman}%
\affiliation{ Lawrence Livermore National Laboratory, Livermore, California 94550, USA}%
\author{S. A. MacLaren}%
\affiliation{ Lawrence Livermore National Laboratory, Livermore, California 94550, USA}%
\author{C. R. Weber}%
\affiliation{ Lawrence Livermore National Laboratory, Livermore, California 94550, USA}%
\author{D. D.-M. Ho}%
\affiliation{ Lawrence Livermore National Laboratory, Livermore, California 94550, USA}%
\author{L. S. Leal}%
\affiliation{ Lawrence Livermore National Laboratory, Livermore, California 94550, USA}%
\author{C. A. Walsh}%
\affiliation{ Lawrence Livermore National Laboratory, Livermore, California 94550, USA}%
\author{J. D. Moody}%
\affiliation{ Lawrence Livermore National Laboratory, Livermore, California 94550, USA}%


\date{\today}

\begin{abstract}
Motivated by breakthroughs in inertial confinement fusion (ICF), first achieving ignition conditions in National Ignition Facility (NIF) shot N210808 and then laser energy breakeven in N221204,\cite{abu1,abu2} modeling efforts here investigate the effect of imposed magnetic fields on integrated hohlraum simulations of igniting systems. Previous NIF experiments have shown fusion yield and hotspot temperature to increase in magnetized, gas-filled capsules. \cite{moody1} In this work, we use the 2D radiation-magnetohydrodynamics code \lasnex{}\cite{zimmerman75} with a Livermore ICF common model.\cite{strozzi1} Simulations are tuned to closely approximate data from unmagnetized experiments. Investigated here is the effect of imposed axial fields of up to 100 T on the fusion output of high-performing ICF shots, specifically the record BigFoot shot N180128, and HYBRID-E shots N210808 and N221204.  The main observed effect is an increase in the hotspot temperature due to magnetic insulation. Namely, electron heat flow is constrained perpendicular to the magnetic field and alpha trajectories transition to gyro-orbits, enhancing energy deposition.\cite{perkins2} In addition, we investigate the impact of applied magnetic fields to future NIF designs, specifically an example Enhanced Yield Capability design with 3 MJ of laser energy and a high-$\rho R$, low implosion velocity Pushered-Single-Shell design. In conclusion, magnetization with field strengths of 5-75 T is found to increase the burn-averaged ion temperature by 50\% and the neutron yield by 2-12$\times$. Specifically, we see yield enhancement of at least 50\% with a 5-10 T applied magnetic field for N221204, while a 65 T field on N210808 with drive symmetrization gives an 8$\times$ increase in yield. 
\end{abstract}

\maketitle



\section{Introduction}

Magnetized ICF is a relatively old concept, dating back at least to the 1980's in established literature.\cite{lindemuth,jones,guskov} The basic promise is that of boosted ignition conditions and yields via reduced conduction losses and enhanced alpha energy deposition into the ignited hotspot due to enhanced magnetic field strengths, where Tesla-scale applied magnetic fields are compressed by the implosion. In Ref. \onlinecite{lindemuth}, reduced order models informed a relatively large parameter scan suggesting magnetization, starting from an already compressed capsule, could enable designs with less stringent fuel densities, lower implosion velocities, and lower drive requirements that also have significant fuel burn-up, a motivation that continues in recent studies.\cite{ho1} Ref. \onlinecite{jones} suggests that magnetization can enhance the yield for volumetric burn, particularly when burn propagation does not inherently occur but may be detrimental in robustly burning circumstances. This is the origin of the concept that magnetic fields may inhibit burn propagation in igniting targets. One of the primary conclusions of the current study is that applied magnetic fields do not inherently degrade igniting targets and, in fact, may more than double the yield of such designs.

Experiments have subsequently validated several of the basic ideas of these early modeling studies, with the notable exception that no magnetized, cryogenic implosions with ice layers have been conducted thus far. Cylindrical and spherical implosions with direct laser drive at the OMEGA laser facility \cite{knauer1,chang,hohenberger} demonstrated field compression and magnetized hotspot insulation according to radiation-magnetohydrodynamic (rad-MHD) modeling.  Pulsed-power-driven implosions of magnetized cylindrical liners at the Sandia Z Machine provide similar validation in a different regime, relevant to the MagLIF fusion concept.\cite{slutz,gomez}  More recent hohlraum-driven NIF experiments \cite{moody1, sio1, strozzi1} with an imposed initial field up to 28 T have shown yield increases up to 2.9x and burn-averaged $T_{ion}$ increases up to 1.4x, consistent with 2D \lasnex{} rad-MHD modeling.\cite{strozzi1}  These experiments used $D_2$ gas-filled capsules and $\lesssim$ 1 MJ of laser energy, due to current practical constraints that magnetized NIF shots must be done at room temperature and with low enough laser energy to avoid risk of optics damage from Brillouin backscatter.  Magnetized NIF experiments also examine other topics, like laser-heated gas pipes relevant to MagLIF\cite{pollock} and high temperature X-ray sources.\cite{kemp,leal} 

This study is the first to present state-of-the-art, fully-integrated rad-MHD modeling of hohlraum-driven, ignition-class implosions with an imposed magnetic field. Where early modeling studies were restricted to reduced-order rad-MHD calculations with strictly azimuthal fields or capsule-only simulations with a frequency-dependent-source (FDS) drive, we are now able to rigorously model magnetized ICF at an appreciable level of fidelity.  Modern rad-MHD codes can model full 2D and 3D capsule and hohlraum systems with expanded, non-ideal MHD. These have found that, with the application of more realistic axial fields, initially $\sim$10 T fields can be compressed to $\sim$10 kT levels to enhance ICF designs, due to factors such as increased robustness to surface roughness, low-mode shape, and residual kinetic energy.\cite{perkins1, perkins2, strozzi2, ho2}  3D modeling with the Gorgon code of the planar magnetized Rayleigh-Taylor instability has shown reduced growth for modes with k-vectors parallel to the field,\cite{walsh1} though magnetized instability modeling of a spherical, integrated implosion has not been published thus far. In addition, there is work looking into the effect of magnetization on hot electron dynamics\cite{strozzi2} and burn propagation, where some inhibition of burn across field lines has been observed in ignition-style designs.\cite{oneill}

In this study we model various hohlraum-driven, ignition-class NIF designs and the effect magnetization has on them. It is found that we are able to routinely enhance the yield of full-scale shots like N180128, N210808, and N221204 by at least a factor of $\sim 2\times$ while potentially approaching $8\times$ for N210808 at 65 T. The shot nomenclature $NYYMMDD$ stands for NIF shot $N$, year $20YY$, month $MM$, and day $DD$. There is a notable $1.5-2\times$ increase in all three cases with the application of just a 5-10 T magnetic field. The currently studied Enhanced Yield Capability (EYC) design very robustly ignites,\cite{maclaren1} meaning the benefit of magnetization is significantly reduced at least for the ad-hoc application of an external field. The Pushered Single Shell (PSS) suggests a way to better take advantage of magnetization given its cooler baseline temperature.\cite{maclaren2,maclaren3,dewald,ho3} 

We present this work in two companion papers: Paper I, here, where we survey several ignition-class designs and their response to applied magnetic fields, and a follow-on Paper II where we go into more details regarding the physics of magnetized ignition. Paper I is organized as follows. In Sec. \ref{nifdesigns} we survey the characteristics of the three historically high-performing NIF designs as well as those of two potential future designs (a 3 MJ EYC and a 1.9 MJ PSS) that have been considered with respect to magnetization.  In Sec. \ref{methodology} we go over the basics of our modeling methodology as well as the tuning process used to most accurately model the experiments in consideration. This is elaborated on in Appendix \ref{appendix}. In Sec. \ref{physics} we survey the basics of magnetized hohlraum and capsule physics that contribute to magnetized ICF. Sec. \ref{magicf} considers the effect of an applied axial magnetic field to the three historically high-performing NIF designs on scalar quantities such as yield and \Ti. We also consider drive asymmetry and how this compares to the idealized, quasi-1D, symmetrized results. Next, we consider the effect of magnetization on future designs PSS in Sec. \ref{pss} and EYC in Sec. \ref{eyc}. Lastly, Sec. \ref{conclusion} summarizes the results and charts potential future work with respect to ICF design.


\section{NIF Designs for Magnetization} \label{nifdesigns}

In this study, we focused on five NIF designs for magnetization, three historically high-performing shots, and two potential future designs. The design parameters can be found in Table \ref{tab:tabtarg}. The selection of shots was not arbitrary; they were all high-performing designs either of historical importance or notable promise based on best practices. All targets used a deuterium-tritium (DT) ice fuel layer and a high-density carbon ablator (HDC) doped with tungsten (W), except for PSS, which used a beryllium (Be) ablator doped with molybdenum (Mo) and traces of argon (Ar) from production. The PSS design is even more unique in that at the outer radius it begins with just Be but then smoothly grades up in density using a dopant. $\Delta \lambda$ is the difference in wavelength between the inner (23$^\circ$, 30$^\circ$) and outer (44$^\circ$, 50$^\circ$) laser cones, which is known as two-color tuning. However, for EYC and PSS we used three- and four-color tuning as well, i.e., separate $\Delta \lambda$ for 23-30$^\circ$ and 44-50$^\circ$ quad pairs, to help with shape and reduce potential stimulated Brilloiun scattering (SBS). In this paper we will define $\Delta \lambda$ in terms of a triplet, where the first number defines two-color wavelength difference (inners versus outers), the second the three-color (23-30$^\circ$), and the third the four-color (44-50$^\circ$), i.e., $\Delta\lambda_{210808}=(1.8,0,0) \ \text{A}^\circ$ but $\Delta\lambda_\text{EYC}=(3.5,1,1) \ \text{A}^\circ$. It is acknowledged that there are many other designs that could be explored, the potential parameter space is enormous. Alternative approaches that could be integrated into a magnetized ICF design were not considered, such as plastic ablators, fast and shock ignition,\cite{strozzi3} and wetted foams, for example. 

The first shot in consideration was N180128. This record BigFoot shot achieved a yield of $Y=40$ kJ and $T_{\text{ion}}=4.9$ keV \cite{baker} and was the record yield at the time. Besides its historical importance, one intention for investigating this design was to see if magnetization could take a high-performing, non-igniting design and push it over the so-called ignition cliff. This was not found to be the case for N180128 with no other changes but may be possible with further optimization.\cite{walsh4} The second was N210808,\cite{abu1} a HYBRID-E (High Yield Big Radius Implosion Design, the E-th iteration) design, which was the first to achieve greater than 1 MJ yield and exceed all extant Lawson criteria for ICF ($Y=1.37$ MJ, \Ti$=10.8$ keV). The third case example was N221204,\cite{abu2} the first shot to achieve energy breakeven, where more energy was generated than delivered to the hohlraum ($Y=3.2$ MJ, \Ti$=13.1$ keV). There have been many minor, cumulative modifications in designs with each shot, but the most pronounced change had been in the increase in laser energy ($E_{L,180128}=1.801$ MJ, $E_{L,210808}=1.890$ MJ, $E_{L,221204}=2.050$ MJ), and notably the greater radiation temperature efficiency of the Hybrid-E hohlraum. 

Two future designs were considered as well. Specifically, a preliminary EYC design,\cite{maclaren1} that extends standard HYBRID-E designs from 2 MJ to 3 MJ of laser energy with a peak power of 450 TW. Simple hydroscaling of a design like N221204 is not an option due to laser optics damage exacerbated by higher laser powers. Therefore, new design parameter spaces must be explored. The second is the PSS design driven by 1.9 MJ of laser energy,\cite{maclaren2,dewald,ho3} which is a platform tailored to radiochemistry studies and uses a density gradient of Mo-doped Be as the ablator instead of CH or HDC. This design is perhaps ideal for magnetization because it inherently has a slower, high $\rho R$, low-temperature implosion that can be directly enhanced by the application of an external magnetic field. 

\begin{figure}
\includegraphics[width=\columnwidth]{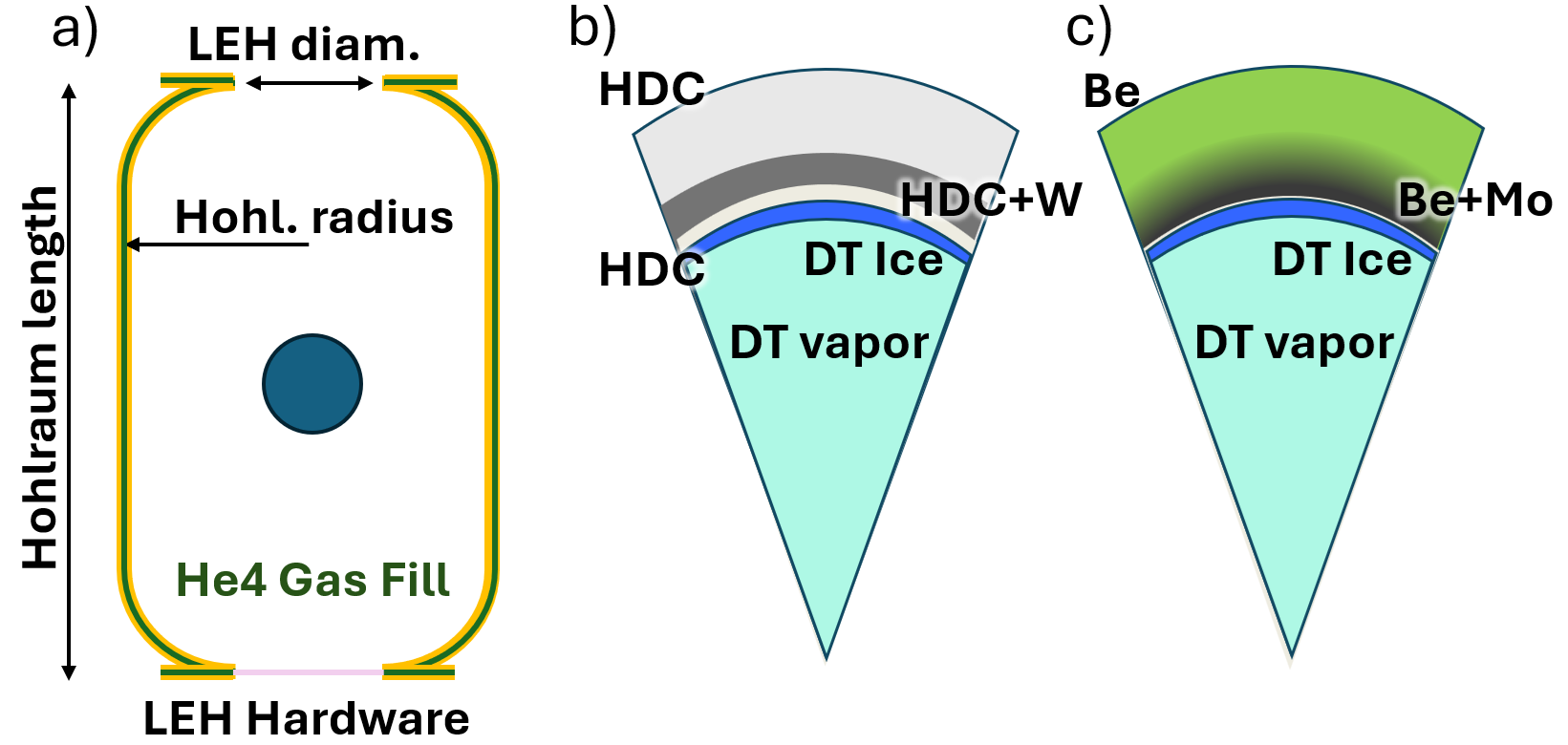}
\caption{\label{fig:targets} Example visualizations of the (a) hohlraum, (b) HDC capsule, and (c) PSS capsule parametrization used in this study.}
\end{figure}

\begin{figure*}
\includegraphics[width=\textwidth]{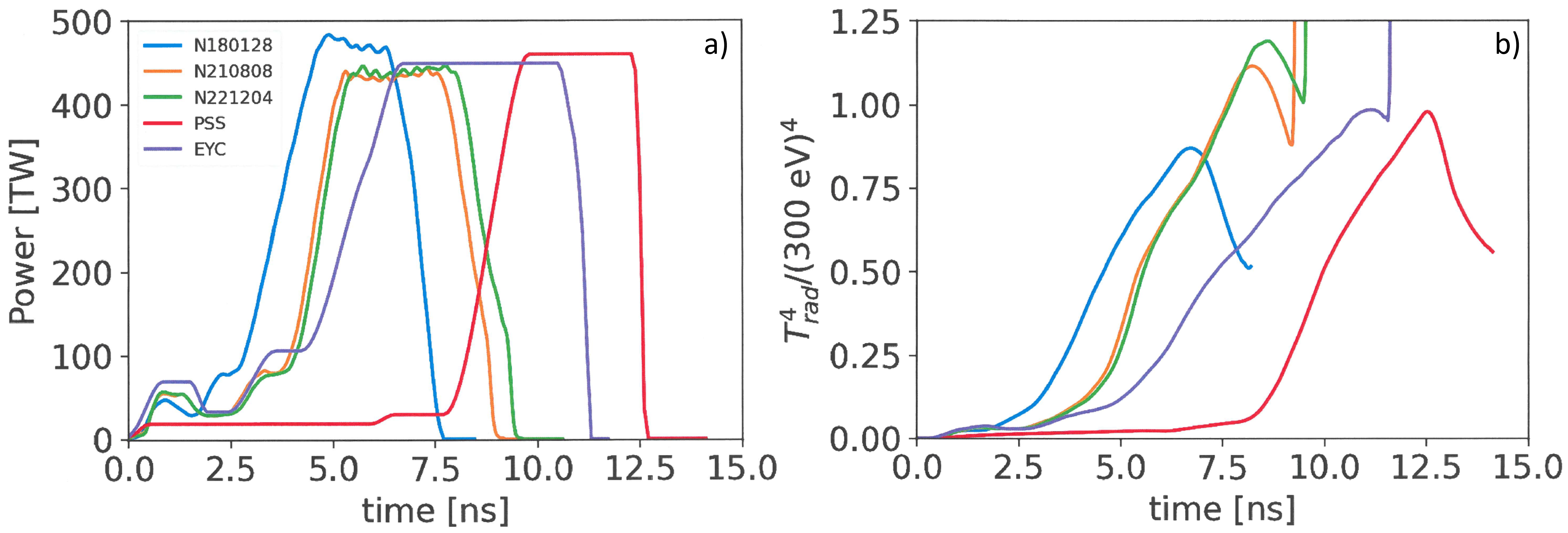}
\caption{\label{fig:laserpower} Comparison of (a) the laser drive profiles  and (b) the corresponding $T_\text{rad}$ profiles predicted by the simulations for N180128 (blue), N210808 (orange), N221204 (green), EYC (purple), and PSS (red). N180128 is a BigFoot design while N210808, N221204, and EYC are all HYBRID-E. The rapid rise in $T_{rad}$ late in time for the HYBRID-E shots is posited to be due to reheating of the hohlraum wall by the high yields generated. This reheating was observed in N221204 but the appropriate diagnostic was not available for N210808.}
\end{figure*}

In Fig. \ref{fig:targets} we show a depiction of the hohlraum and target capsules relative to the experimental properties in Table \ref{tab:tabtarg}. Four of the shots had qualitatively similar HDC capsule designs, while PSS has a unique design. A comparison of the laser power profiles and corresponding hohlraum radiation temperature $T_{rad}$ time history can be seen in Figs. \ref{fig:laserpower} (a) and (b), where $T_{rad}$ is the average radiation temperature across a spherical annulus outside the ablation front of the capsule. N180128 has a distinct BigFoot design while N210808, N221204, and EYC all have HYBRID-E designs. PSS has a distinctly longer laser foot because Be does not need a picket, which is deleterious to symmetry control, due its much lower melt pressure that keeps the 1st shock rarefaction  pressure above re-freeze and better ablation front stability.\cite{zylstra2018} HDC on the other hand uses a picket due to the higher melt temperature. As is apparent, HYBRID-E designs presented a more efficient approach than BigFoot while EYC has an extended peak to take advantage of greater energy  without the higher peak power dangerous to the optics. In Fig. \ref{fig:laserpower} (b) we have raised $T_{rad}$ to the fourth power and normalized to a standard value of (300 eV)$^4$ to represent x-ray flux $M \propto T_{rad}^4$. Uniquely, the high yield shots of N210808, N221204, and EYC all demonstrate signs of intense $T_{rad}$ peaks at or after bangtime. It is posited that this is hohlraum wall reheating,\cite{rubery} where blast-wave-driven detritus from the ablator shell in turn heats the inner surface of the gold-plated, depleted uranium (Au+DU) hohlraum walls. In Table \ref{tab:table1} we present a comparison of the metrics for the experimental shots (N180128, N210808, and N221204) relative to our baseline simulations at $B=0$  T. As discussed in Appendix \ref{appendix}, the primary goal here was to match closely bangtime and $p_2/p_0$ hotspot shape and then to get yield and \Ti as close as possible, where $p_N$ is the $N$-th Legendre polynomial and \Ti is the burn-weighted ion temperature over space and time. Per standard practices in the ICF community these tunes were quite close, with an exception for N221204 which is about 50\% too high in yield. This was deemed acceptable given the rapid rate of performance increase in recent experiments based on the N221204 design. Other metrics of interest are included, such as peak implosion velocity $V_\text{imp}$, primary neutron down-scattered ratio (DSR), a figure of merit that is directly proportional to the fuel areal density $\rho R$ at stagnation,\cite{knauer2} peak radiation temperature $T_\text{rad}$, and the burn-up fraction of the DT fuel.

\begin{table*}
\caption{\label{tab:tabtarg} Hohlraum, capsule, and laser design parameters for the high-performing NIF shots in question, N180128, N210808, and N221204, in addition to future designs for the EYC and PSS. These are based on the latest metrics used in post-shot modeling and may differ slightly from previously published results. $\Delta\lambda$ is based on either the experimental result or the preshot design, where $a/b/c$ corresponds to the $\Delta\lambda$ between the a) inner (23$^\circ$, 30$^\circ$) and outer (44$^\circ$, 50$^\circ$) beams, b) 23$^\circ$ and 30$^\circ$, and c) 44$^\circ$ and 50$^\circ$.}
\begin{ruledtabular}
\begin{tabular}{l|c|c|c|c|c}
Shot & {BigFoot}   & {Lawson} & {Breakeven} & Enhanced Yield & Pushered Single\\
     & {(N180128)}  & {(N210808)} & {(N221204)} & Capability (EYC) & Shell (PSS) \\
\hline
LEH diam. [mm]  & 3.9  &  3.1 &  3.1 & 4.15 & 3.37  \\
Hohl. rad. [mm]  &  3.0  & 3.2  & 3.2  & 3.52 & 3.2\\
Hohl. len. [mm]  &  11.31 &  11.26 & 11.25 & 13.93 & 11.26 \\
Hohlraum material  & Au  &  \multicolumn{4}{c}{Au+DU}  \\
He4 fill [mg/cm$^{3}$]  &  0.298  & 0.301  & 0.297  & 0.300 & 0.300 \\
  &   &   &  &  \\
\hline

Outer rad. [$\mu$m]  & 1021.7 &  1128.4 &  1135.5 & 1312.0 & 1175.0 \\
Abl. mat.  &  \multicolumn{4}{c|}{HDC} & Be+Ar \\
Abl. thick. [$\mu$m] & 71.10  &  79.35 &  84.13 & 82.75 & 205 \\
Abl. den. [g/cm$^{3}$]  &  3.44 & 3.35  &  3.35 & 3.35 & 1.8-1.944 \\
$f_\text{dopant}$ [at. \%]  &  W:0.27  &  W:0.45 & W: 0.63  & W: 0.45 & Mo: 0.0-21 \\
Ice thick. [$\mu$m]  &  49.0 &  65.9  &  63.7 & 69.2 & 50 \\
Ice den. [mg/cm$^{3}$]  & \multicolumn{5}{c}{0.25}  \\
DT vapor [mg/cm$^{3}$]  &  0.45 &  0.44 &  0.48 & 0.44 & 0.44 \\
Vapor frac. [at. \%]  & \multicolumn{5}{c}{D-60\%, T-35\%, He$^3$-4\%, H-1\%}  \\
  &   &   &  &  \\
\hline
Laser energy [MJ]  &  1.8 &  1.89 & 2.05 & 3.0 & 1.9 \\
Peak power [TW]  &  480 & 441  &  440 & 450 & 450\\
$\Delta\lambda$ [$A^\circ$]  &  0/0/0 & 1.8/0/0  &  2.75/0/0  & 3.5/1/1 & 4.5/0/0.4 \\
\end{tabular}
\end{ruledtabular}
\end{table*}

\begin{table}
\caption{\label{tab:table1}Performance metrics for the high-performing NIF shots in question, N180128, N210808, and N221204, for the experiments themselves and the final result of the simulation tuned to match them.}
\begin{ruledtabular}
\begin{tabular}{l|cc|cc|cc}
Shot & \multicolumn{2}{c|}{BigFoot}   & \multicolumn{2}{c|}{Lawson} & \multicolumn{2}{c|}{Breakeven} \\
     & \multicolumn{2}{c|}{(N180128)}  & \multicolumn{2}{c|}{(N210808)} & \multicolumn{2}{c|}{(N221204)} \\
\hline
    & exp\cite{baker} & sim & exp\cite{abu1} & sim & exp\cite{abu2} & sim  \\
\hline
Yield [MJ] & 0.05 & 0.047 & 1.37  & 1.31 & 3.2 & 4.6 \\
Bangtime [ns] & 8.01 & 7.99  & 9.27 & 9.29 & 9.53 & 9.56 \\
\Ti [keV]   & 4.9 & 4.44 & 10.9 & 9.6  & 13.1  & 11.8 \\
p2/p0 [\%] & 1.7 & 0.5 & -6 & -1.2  & -0.4 & -0.1 \\
p4/p0 [\%] & -0.01 & -0.3 & 14 & 0.3  &  n/a & 8.4 \\
p6/p0 [\%] & n/a & -3.9 & n/a & -7.3  & n/a  & 1.3 \\
DSR [\%] & 3.05 & 3.25 & 2.75 & 3.20  &  3.01 & 2.98 \\
$T_\text{rad}$ Peak [eV]  & 284  & 290 & 307  & 308 & 313  & 314 \\
Burnup frac. [\%] & 0.12 & 0.11 & 1.77 & 1.84  &  4.33 & 6.04  \\
\end{tabular}
\end{ruledtabular}
\end{table}

\section{Simulation Methodology and Tuning} \label{methodology}

All simulations performed in this study were done with the \lasnex{} rad-MHD code \cite{zimmerman75} and in particular via the \lasnex{} Hohlraum Template (LHT).\cite{strozzi1} The LHT is part of an effort to unify standards and physics packages across ICF modeling with \lasnex{}. All \lasnex{} simulations were done in RZ-2D geometry and included both the hohlraum and capsule, known as a fully integrated simulation. \lasnex{} uses 3D ray tracing to propagate lasers through the hohlraum with refraction and inverse bremsstrahlung absorption with corrections due to the Langdon effect.\cite{langdon1,sherlock1} NIF has four cones of beams at 23$^{\circ}$,  30$^{\circ}$, 44$^{\circ}$, and 50$^{\circ}$ from the vertical axis. Cross-beam energy transfer (CBET) via ion acoustic waves can transfer energy to the laser with the lower frequency in the plasma flow frame and this is modeled with an inline CBET package in \lasnex{}.\cite{kruer1,michel1,higginson1,strozzi2017}  

\begin{figure}
\includegraphics[width=0.9\columnwidth]{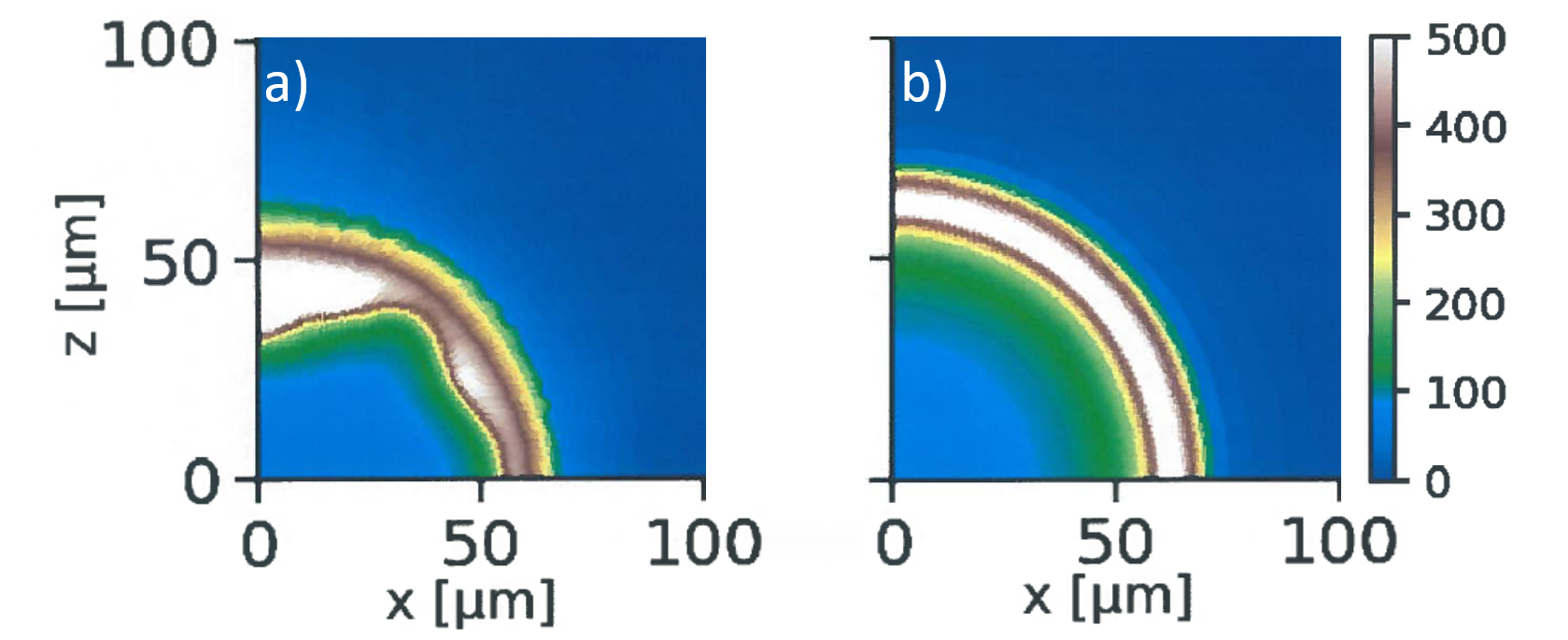}
\caption{\label{fig:symcomp} Plot of density $\rho$ [g/cm$^{3}$] for N210808 when (a) unsymmetrized and (b) symmetrized. Polar caps as seen in a) are standard features in current ICF designs, i.e., HYBRID-E.}
\end{figure}

The ideal implosion in 3D without magnetization is a perfectly spherical implosion, which compresses with $r^3$ scaling. Asymmetric degradations and external magnetization can push the implosion towards an $r^2$ scaling, like with cylindrical compression. Magnetization is an inherently asymmetric effect, but we can consider the spherical case as a point of optimal performance and see the effect of MHD under ideal conditions. We do this in \lasnex{} by fully averaging the x-ray coupling to electrons over angle in the capsule. This method is called symmetrization and symmetrized implosions give us an idea of the ideal 1D result while still considering higher-order, multi-dimensional effects like MHD. We typically do this without CBET, which makes results almost entirely reproducible and more computationally affordable. A comparison of an implosion when run normally versus when symmetrized can be seen in Fig. \ref{fig:symcomp}, where the symmetrized run is effectively equivalent to a 1D implosion, which is the ideal result for ICF. However, in reality x-ray symmetry swings are inherent to hohlraums, given that the walls and Laser Entrance Holes (LEHs) evolve in time.  A purely symmetric drive throughout time is impossible, even if one can achieve symmetry "swings" that produce an implosion that is round with respect to some integrated metric. Even if someone minimizes the $p_2/p_0$ shape of an implosion with swings, there will still be internal flows and residual kinetic energy present that are driven by time-dependent asymmetries. 

We model non-local thermodynamic equilibrium (NLTE) atomic physics with the Detailed Configuration Accounting (DCA) suite,\cite{scott,scott2022} for material properties including opacity and equation of state in hohlraum regions with electron temperatures above 300 eV. We include external LEH hardware, such as an aluminum washer,\cite{higginson1} which can affect LEH conditions and CBET. \lasnex{} also has a robust multi-ion-species hydrodynamics package based on a 13-moment formulation in Ref. \onlinecite{schunk} and extended MHD that is necessary for this study and generally elevates the fidelity of modeling ICF physics. A more detailed explanation of the MHD package can be found in Ref. \onlinecite{strozzi1}. In addition, there is a buoyancy-drag, annular shell mix model \cite{zhou} that we use to degrade simulation yields approximately in line with experimental observations.\cite{bachmann} This is intended to include the effects of hydrodynamic mix, e.g., due to the Rayleigh-Taylor instability and capsule fill-tube, in simulations that do not resolve this physics. We currently do not have an effect of magnetization on such mix inside our models. 

ICF simulations typically overpredict the hohlraum x-ray drive and thereby under-predict capsule bangtime. A common solution to this issue is to use a power multiplier on the drive laser, typically around 0.8 at peak power and additional multipliers in the foot to match shock timings. Wall opacity multipliers are seen as a more favorable solution, given the higher uncertainty in the NLTE physics versus laser deliver measurements.\cite{divol} Lower power results in different laser absorption and lower \Te which modifies CBET, but in this study we only used power multipliers since that is a more mature method and the object of our study is not hohlraum physics but rather the implosion itself. We also use tuning parameters for CBET and atomic mix to match implosion shape and fusion yield.

To achieve the best results for matching the simulations to the experiments in question, we go through a multi-step tuning process. In this study, we used the Automated NIF Tuning Suite (ANTS)\cite{weber} framework for tuning rad-hydro simulations to experiments that utilizes Gaussian-process-based Bayesian optimization. Here, we went through three steps:
\begin{enumerate}
  \item Power multipliers of foot $(P_n)$ to match shock timing data of VISAR from an associated keyhole experiment, where $n$ corresponds to a point in time along the laser based on a time-dependent, spline-based decomposition of the laser determined by ANTS.
  \item Peak power $(P_\text{peak})$, CBET clamp (the ion-wave electron density fluctuation $\delta n/n$), and laser cone fraction $(CF)$ multipliers to match bangtime and $p_2/p_0$ shape of the 13-15 MeV neutron profile of the primary DT experiment. The CBET clamp defaults to 0.005 before being changed later in time after the laser foot, while the CF multiplier when used is constant throughout.
  \item Three mix model parameters to match yield $(Y)$, $p_2/p_0$, and minimize or match $p_4/p_0$ and $p_6/p_0$ if data is available for the primary DT experiment.
\end{enumerate}
A detailed description of the tuning process can be found in Appendix \ref{appendix} as well as Ref. \onlinecite{schmit}.

\section{Basic Physics of Magnetized Hohlraums} \label{physics}

The application of 10-100 T magnetic fields onto an ICF target is a non-trivial physical and engineering problem. For example, magnetized ICF experiments on NIF necessitated the development of a new gold-tantalum alloy hohlraum which has comparable X-ray generation as pure gold but an increased electrical resistivity so that it is not crushed by the applied field.\cite{moody1} For a magnetic field to have a perturbative effect on an ICF hotspot plasma, kT-scale fields are necessary. However, this is actually a relatively simple goal to achieve given the properties of an ICF implosion. An ICF capsule, when ideally compressed, will have a convergence ratio of $\sim$30$\times$, i.e., initial/final capsule radius, which means that the area of a magnetic flux surface that spans the diameter of the capsule will be compressed by a factor of ~900$\times$.
Therefore, an applied 30 T field, which has been almost achieved in the NIF subscale, symcap experiments thus far,\cite{sio1,moody1,strozzi1} will ideally reach 27 kT scales at stagnation. This field is large enough to have an impact even with some flux loss, e.g., through resistive diffusion and the Nernst effect.

Flux compression can be described by Alfv\'{e}n's theorem, which says that magnetic flux in a conducting fluid is conserved through deformations of material surfaces, i.e., $D\Phi_B/Dt=0$, where $\Phi_B$ is the magnetic flux and $D/Dt$ is the advective or material derivative.\cite{blackman} In an ideal plasma that undergoes a reduction in radius from $r_0$ to $r_1$ and where magnetic flux remains constant, $\Phi_0=A_0 B_0 = \Phi_1=A_1 B_1$, where $A$ is the surface area in question, the magnetic field scales as: 
\begin{equation}
B_1  = \left ( \frac{r_0}{r_1} \right )^2 B_0.  
\end{equation}
This process is governed by the induction equation (here simplified versus the one implemented in \lasnex{}): 
\begin{equation}
\label{induction}
\partial_t\myvec{B}=\eta\nabla^2\myvec{B} + \nabla\times[(\myvec{v} + \myvec{v}_N)\times\myvec{B}], 
\end{equation}
\begin{equation}
\label{nernstspeed}
\myvec{v}_N \equiv -\frac{\beta_\wedge}{e |B|}\nabla T_e,
\end{equation}
 where $\myvec{v}_N$ is the Nernst velocity, $\beta_\wedge$ is a Braginskii transport coefficient, and $T_e$ is the electron temperature. The Nernst effect transports perpendicular field lines down $T_e$ gradients.
 Ideal MHD applies to magnetic induction when diffusion and Nernst advection are negligible: $\eta, \myvec{v}_N \rightarrow 0$. 

Extremely high, kT fields in the capsule near stagnation bring the two primary benefits to ICF, magnetoinsulation and instability reduction. Magnetoinsulation primarily takes advantage of the reduced electron heat flux $\textbf{q}=\kappa\cdot\nabla T$ perpendicular to magnetic field lines. 
We can rewrite the heat flux in terms of $\boldsymbol{\kappa}$, the anisotropic thermal conductivity tensor, which gives us
\begin{equation}
\textbf{q} = -\kappa_\parallel \hat{\textbf{b}}(\hat{\textbf{b}}\cdot\nabla T) - \kappa_\perp [\nabla T-\hat{\textbf{b}}(\hat{\textbf{b}}\cdot\nabla T)] - \kappa_\wedge \hat{\textbf{b}}\times \nabla T,
\label{anisotropic}
\end{equation}
where $\hat{\textbf{b}}$ is the unit vector along \textbf{B}, and $\kappa_\wedge$ is the Hall (Righi-Leduc) conductivity giving a heat flux perpendicular to both $\hat{\textbf{b}}$ and $\nabla T$. The thermal conductivities in Eqn. \ref{anisotropic} scale as follows:\cite{braginskii1965}
$$ \kappa_\parallel \sim T_e^{5/2},  \kappa_\perp \sim \frac{\kappa_\parallel}{1+(\omega_{ce}\tau_{ei})^2}, \kappa_\wedge \sim \frac{\kappa_\parallel \omega_{ce}\tau_{ei}}{1+(\omega_{ce}\tau_{ei})^2}.$$
These transport properties can be characterized by two dimensionless parameters, the Hall parameter $H_e=\omega_{ce}\tau_{ei}$, where $\omega_{ce}=eB/m_e$ is the electron cyclotron frequency and $\tau_{ei}$ is the electron-ion collision time, and plasma  $\beta=P/P_B=2\mu_0P/B^2$, where $P$ is the thermal pressure and $P_B = B^2/2\mu_0$ is the magnetic pressure. In highly magnetized plasmas we have $H_e>1$, even exceeding 100 within the hotspot at bangtime. With respect to transport properties we can write an effective conductivity as $\kappa_{\text{eff}}\approx\frac{1}{3}\kappa_\parallel\left(1 + 2 \kappa_\perp/\kappa_\parallel \right)$, where for highly magnetized conditions $\kappa_\perp/\kappa_\parallel$ approaches zero as $H_e\gg 1$. However, unlike an underdense plasma found in a Tokamak or outer space, magnetized ICF plasmas have $\beta\gg1$ near stagnation. However, this is not true prior to shock propagation through the gas fill. Magnetic pressure modifies shock propagation and thereby shape prior to stagnation, where shocks propagate faster on the equator versus the poles. In addition, magnetic fields affect alpha energy deposition. The effective stopping length of alphas is decreased as they undergo gyro orbits, defined by $r_g=mv_\perp/|q|B$, as they propagate through the hotspot and into the dense fuel. Additional contributions of extended MHD such as the Righi-Leduc and Biermann battery effects will be discussed in further detail in Paper II. 

The second benefit of magnetic fields is that they are believed to reduce Rayleigh-Taylor instabilities. The basic concept here is that field lines resist perturbations perpendicular to them, i.e., there is a restoring force against the bending of field lines, $\vec{f}_T = (\vec{B}\cdot\nabla)\vec{B}/\mu_0$. This has yet to be demonstrated in integrated ICF experiments but has been seen in simulations.\cite{perkins2,walsh2} The idea that hydrodynamic mix can be reduced via magnetization comes from the concept of magnetic tension, which acts to straighten out bent field lines. The actual dynamics of the evolution of the magnetic field inside the capsule is complicated by various MHD effects, such as the Nernst effect that transports perpendicular field lines down temperature gradients per equations \ref{induction}-\ref{nernstspeed}, i.e., on the equator. However, the primary perturbation to spherical field compression is that of hydrodynamic ablation, which pinches in field lines further inward on the equator as opposed to the poles as the ice layer heats up and ablates into the hotspot. 

\begin{figure*}
\includegraphics[width=\textwidth]{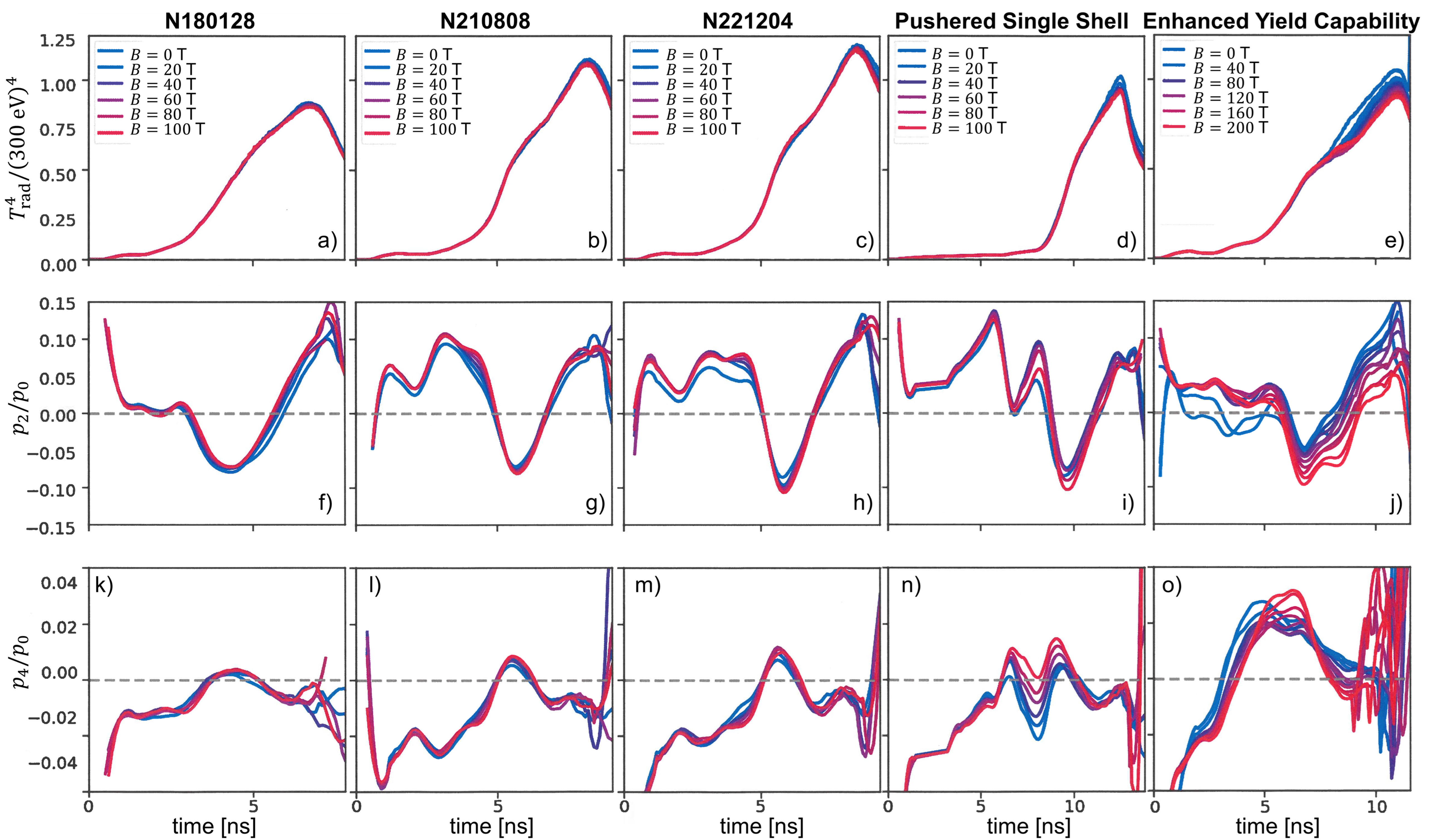}
\caption{\label{fig:magsym} A comparison of the $T_{rad}^4 \propto p_0$ in (a)-(e), $p_2/p_0$ in (f)-(j), and $p_4/p_0$ in (k)-(o) Legendre modes of the x-ray drive at varying magnetic field strengths for the shots in question.}
\end{figure*}

The primary object of this study is that of the implosion itself. However, in order to implode a capsule we need a hohlraum, and we need to validate that we may treat magnetization of the capsule as being relatively independent of the hohlraum. A recent study of fully-integrated \lasnex{} simulations modeling gas-filled symcaps suggested that magnetization at ~$B=30$ T initially does not significantly affect the physics of the hohlraum,\cite{strozzi1} namely the radiation temperature $T_{rad}$ generated by the laser-heated walls of the hohlraum (there is a modest effect on x-ray flux asymmetry per \lasnex{} modeling). This has also been observed in experiments, where the observed $T_{rad}$ when magnetized at $B=26$ T is almost identical to when unmagnetized.\cite{sio1} We verify this for all igniting designs by looking at $T_{rad}$ as a function of time, which is an effective metric of the x-ray drive experienced by the capsule. $T_{rad}$ is practically identical for the duration of the entire implosion in Fig. \ref{fig:magsym} (a)-(e) for all five designs. The variation after bangtime is typical even when unmagnetized and thereby not a concern. The only exception is for PSS and EYC, where there is a slight decrease in $T_{rad}$ late in time, particularly for EYC, but this is something that can be corrected for in future designs.

In addition,  we can extract a Legendre decomposition of the x-ray flux in the hohlraum to get an idea of the degree of asymmetry present in the simulation, as seen in Fig. \ref{fig:magsym} (f) - (o). The divergence of $p_2/p_0$ when magnetized seems small relative to when unmagnetized and $p_4/p_0$ differences are neglible, but shape control is challenging and variable on its own even when unmagnetized. An applied magnetic field may alter CBET both directly, by changing the basic CBET coupling (currently not included in \lasnex{} or any rad-MHD code we are familiar with), and indirectly, by changing the LEH plasma conditions (included in \lasnex{}, to the extent the MHD model is valid).  For magnetized simulations, we use the CBET clamp tuned to match the relevant unmagnetized data.  Even if this is not accurate, small changes in the imposed wavelength shift should be able to tune magnetized implosion shape similarly to unmagnetized cases. In addition, the evolution of the gold bubble on the hohlraum wall is relatively unaffected. Earlier work showed measurements of the electron temperature in the gold bubble of NIF hohlraums only match \lasnex{} modeling when the heat flow is restricted compared to the unmagnetized Spizter-H\"arm level.\cite{meezan2020} The \lasnex{} MHD model naturally provides this, if the Nernst effect is multiplied by $f_N=0.1$.\cite{rosen2018} We follow this and use $f_N=0.1$ outside the capsule, but 1 inside it.  More recent modeling work with modified gold properties (namely opacity and heat capacity) allows us to keep $f_N=1$ in the hohlraum to agree better with NIF data.\cite{chen2024}



\section{Magnetizing High-Performing Ignition Shots} \label{magicf}

This section examines the effect of magnetizing the three NIF designs that have already been shot in our study. The goal of this study was to focus on high-performing shots and the effect that magnetization has on them. Across the board we can see at least a $2\times$ increase in yield for an optimally applied magnetic field and 50\% increase in \Ti, though we are able to get up to $6\times$ or higher under particular conditions. We will systematically go through each shot and then identify differences between the three that may help inform future designs. Properties of the three shots are given in Table \ref{tab:tabtarg}.

\subsection{N180128 - highest performing BigFoot shot}

The BigFoot platform on NIF was intended to be a more predictible and reproducible platform than the earlier HighFoot platform. Prior to the development of the HYBRID-E platform used in the ignition designs for current NIF implosions, BigFoot was a leading design. In addition to having the record yield at the time, $Y=49$ kJ and \Ti= 4.89 keV,  N180128 was notably round and demonstrated alpha heating but not enough to cause it to ignite. 

\begin{figure*}
\includegraphics[width=\textwidth]{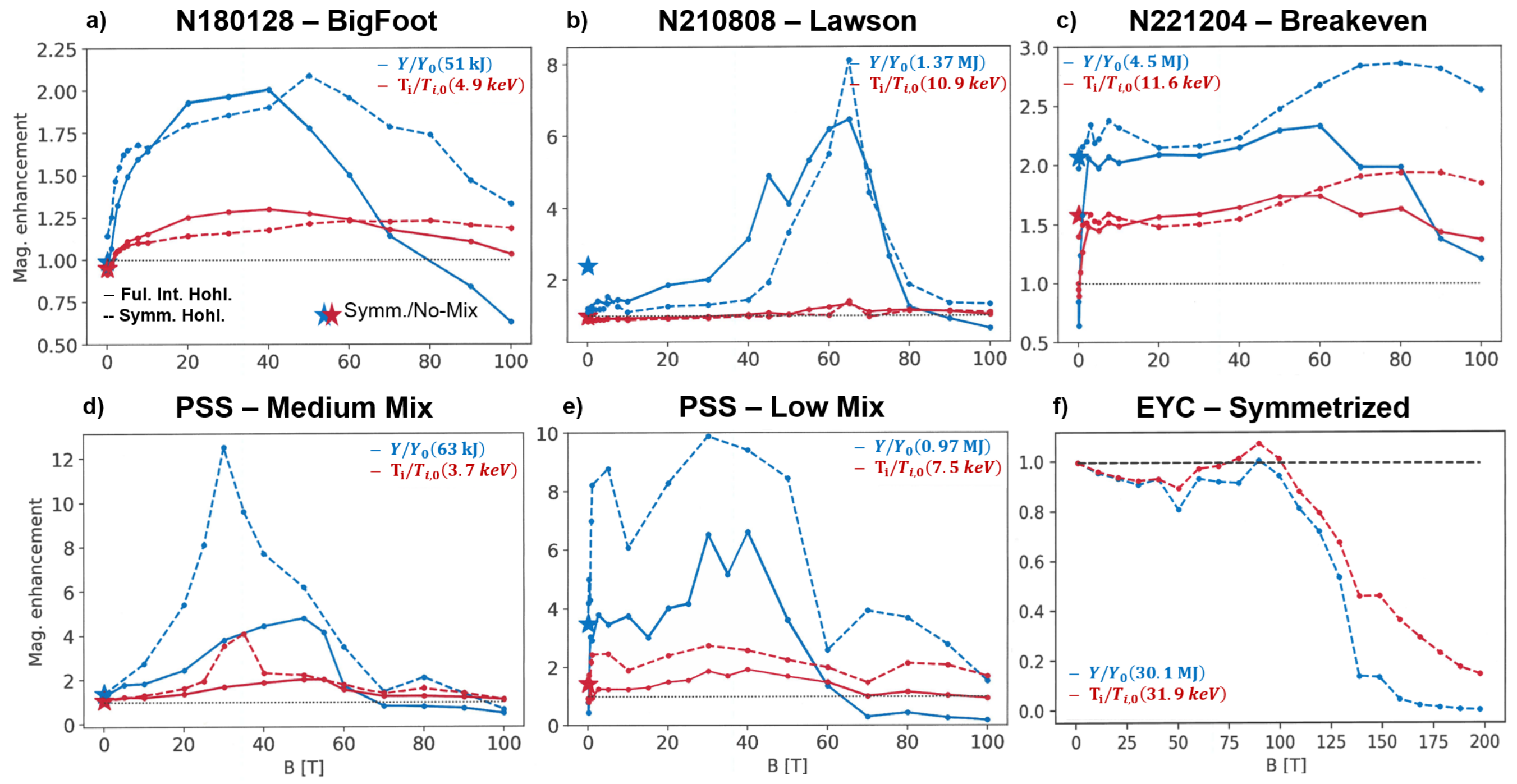}
\caption{\label{fig:bsa} Comparison of yields (blue) and \Ti (red) for the five NIF designs when an external magnetic field is applied (two instances for PSS). In addition, we consider the fully-integrated, asymmetric implosion (solid) and compare it to the symmetrized results (dashed). Stars mark the baseline, unmagnetized result when symmetrized and mix is not enabled, which is effectively the idealized 1D limit. Slight mix can actually enhance performance in certain conditions but is beyond our capabilities to control as of now. Plotted here are results for a) N180128, b) N210808, c) N221204, d) PSS with moderate mix ($\delta h = 0.25 \mu m, f_\text{mix}=0.02 \%$), e) PSS with low mix ($\delta h = 0.2 \mu m, f_\text{mix}=0.015 \%$), and f) EYC. The numbers in the top-right corner of each panel denote the normalization value for each panel with respect to $Y$ and \Ti.}
\end{figure*}

Doing a scan of magnetic field strength we can see the relative enhancement of $Y$ (solid blue) and \Ti (solid red) in Fig. \ref{fig:bsa}(a), where the peak enhancement can be found at $B=30$ T. The enhancement level is normalized to the $B=0$ T value of the simulation scan.
The decline in temperature and yield at higher magnetic field strengths is primarily due to shape degradation from MHD effects. For this case example, the hot spot becomes more oblate when magnetized.
In this case the implosion resisted compression more on the equator as opposed to the poles, due to moderate magnetic pressure effects and reduced heat transport, while the poles allowed for greater transport and dynamic motion. 

As discussed in the introduction, a technique we use is symmetrization, whereby we can mimic the ideal, 1D result in 2D. This can be seen in Fig. \ref{fig:bsa}(a) with the dashed blue and red lines for symmetrized yield and \Ti, respectively. In this case the maximum magnetic enhancment of 2.1$\times$ can be found at 50 T instead of 30 T. 
In addition, when symmetrized, the relative enhancement due to magnetization rises much more rapidly for small, applied magnetic fields. Contrary to expectations, \Ti can actually be lower generally for applied magnetic fields when symmetrized. It has been suggested that for clearly non-igniting implosions, magnetic pressure can be a more prominent effect,\cite{ho2} preventing the implosion from compressing as much, particularly on the equator.

\begin{figure*}
\includegraphics[width=\textwidth]{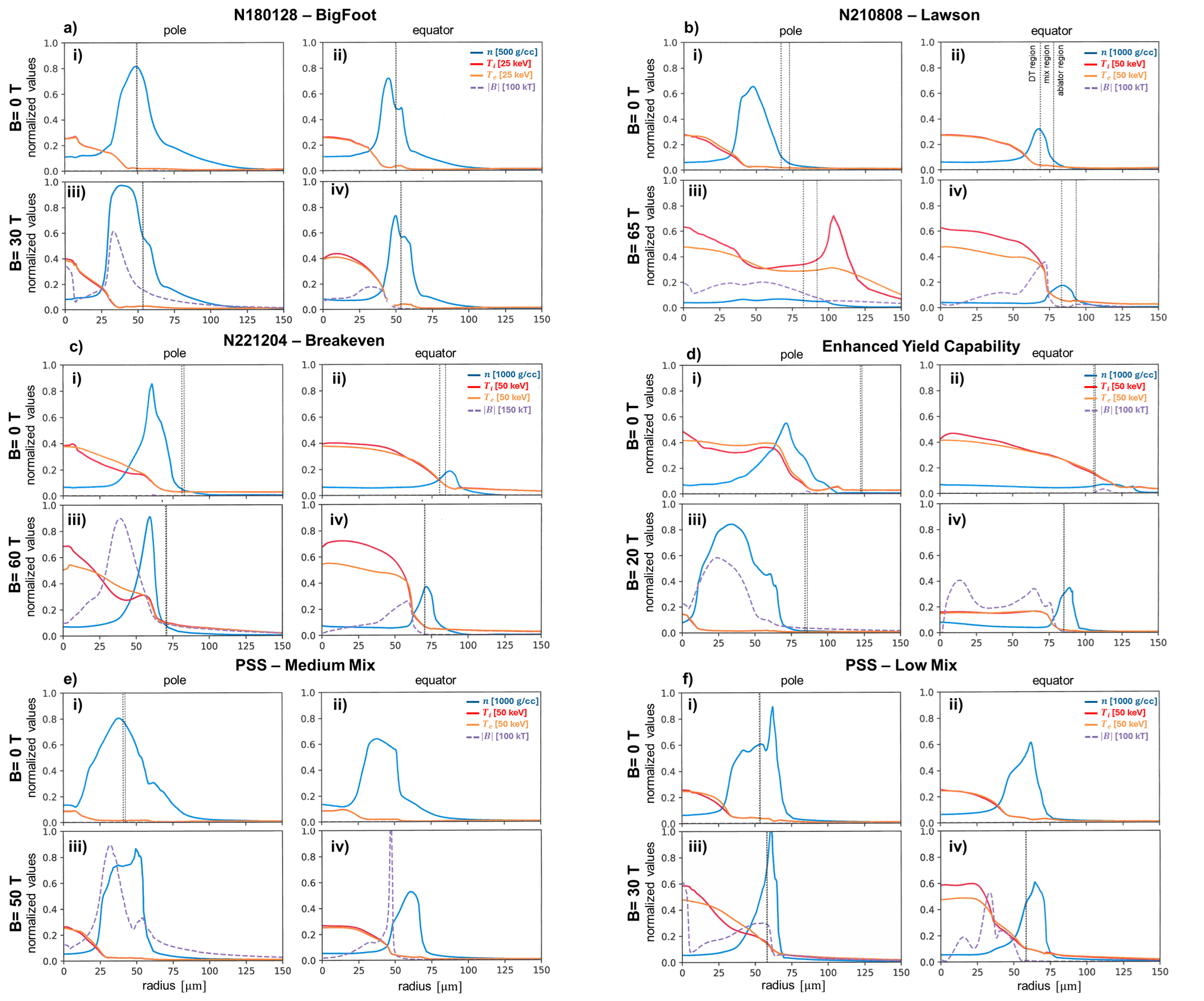}
\caption{\label{fig:los} Comparison of lineouts of key physics parameters along the pole (i),(iii), and the equator (ii), (iv) for a) N180128, b) N210808, c) N221204, d) EYC, e) PSS with medium mix levels, and f) PSS with low mix levels. These lineouts are taken at bangtimes a) unmagnetized $t_{bt,0}=7.99$ ns and magnetized $t_{bt,30}=8.00$ ns, b) $t_{bt,0}=9.26$ ns and $t_{bt,65}=9.25$ ns, c) $t_{bt,0}=9.57$ ns and $t_{bt,60}=9.51$ ns, d) $t_{bt,0}=11.55$ ns and $t_{bt,20}=11.62$ ns, e) $t_{bt,0}=13.90$ ns and $t_{bt,30}=14.00$ ns, f) $t_{bt,0}=13.96$ ns and $t_{bt,40}=14.07$ ns. Each image shows the normalized values for the density $\rho [\text{g/cm}^{3}]$, ion temperature $T_i$ [keV], electron temperature $T_e$ [keV], and magnetic field strength $|B|$ [kT]. Each quad of images is normalized per the values in the legend with respect to the physical quantities plotted. The vertical lines denote the boundary between the DT, mix, and ablator regions, as denoted in b.ii). 
}
\end{figure*}

We can take a closer look at the effect of magnetization on the implosion by taking lineouts of the hotspot at bangtime, as seen in Fig. \ref{fig:los} (a), where we have the unmagnetized case on the pole in panel (i), on the equator (ii), and the magnetized at $B=30$ T, when yield is maximized, on the pole in panel (iii) and the equator in (iv). Plotted are various quantities normalized to different values for clarity. In blue we can see the density profile, where the interface between the ice layer and the ablator is demarcated by the vertical dashed line, and the presence of two vertical lines denotes a mix region. The density profile is almost identical on the equator when magnetized or not but is different on the pole. This may contradict standard assumptions regarding the effects of magnetization, since the Nernst effect and magnetic pressure primarily act perpendicular to magnetic field lines. 

On the polar lineout the magnetized case has a denser, thicker layer than when not magnetized. This is mostly due to a modification of the shape of the implosion as a result of magnetization, i.e., reduced heat transport perpendicular to the fields and modified shock velocities. These effects modify the implosion conditions at bangtime, where magnetization has the greatest effect. \Ti and \Te are depicted in red and orange and are mostly in equilibrium in this case, which is not always the case in this study. When magnetized, \Ti and \Te almost double in the hotspot. This convolved with the local density, which is approximately identical in both cases, results in higher nuclear reaction rates, $f = n^2 \langle \sigma v\rangle$, where the Maxwell-averaged reactivity $\langle \sigma v \rangle$ is a function of \Ti. Lastly, the magnetic field is depicted in dashed purple, where it reaches excesses of 50 kT near the gas-ice interface. The magnetic field is highly dynamic as the implosion reaches stagnation and bangtime so it is difficult to predict its topology. Extended MHD effects contribute to this dynamic but topological changes are primarily driven by hydrodynamic flows and internal ablation of the ice-layer. These cause the magnetic field to distort from an ideal visualization of a compressed bundle of magnetic field lines as might be assumed at first glance.


\begin{figure*}
    \begin{subfigure}{\textwidth}
        \centering
        \includegraphics[width=0.99\textwidth]{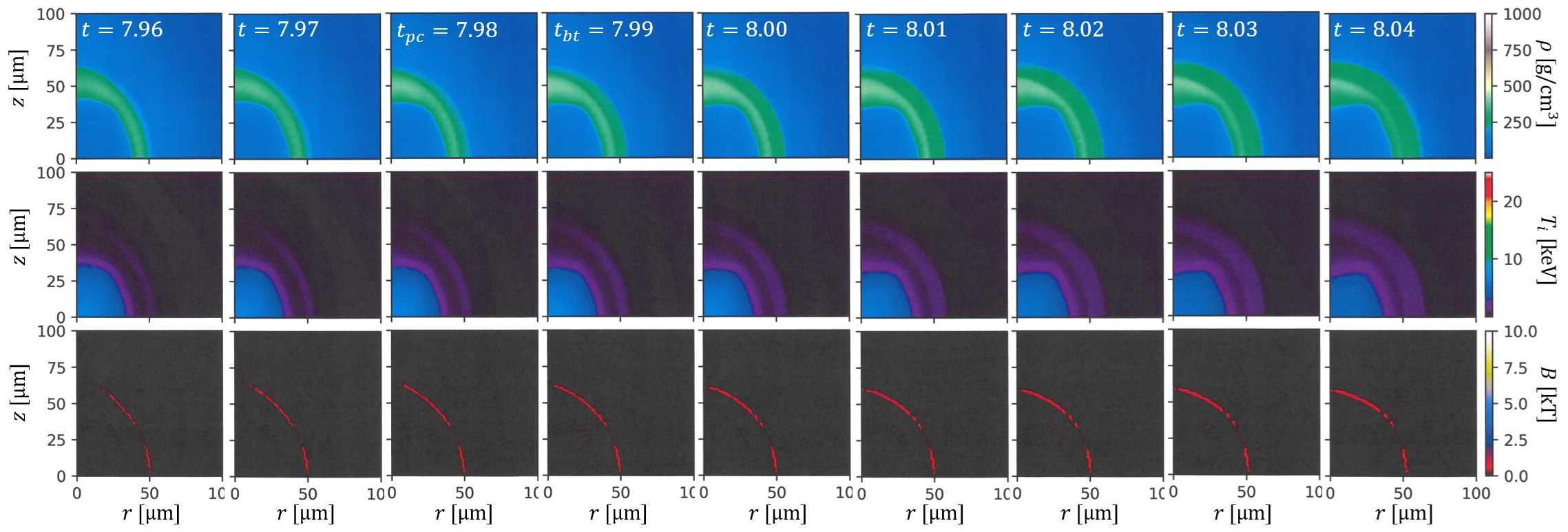}
        \caption{$B=0$, peak compression at 7.98 ns and bangtime at 7.99 ns.}
        \label{fig:sub1}
    \end{subfigure}
    \begin{subfigure}{\linewidth}
        \centering
        \includegraphics[width=0.99\textwidth]{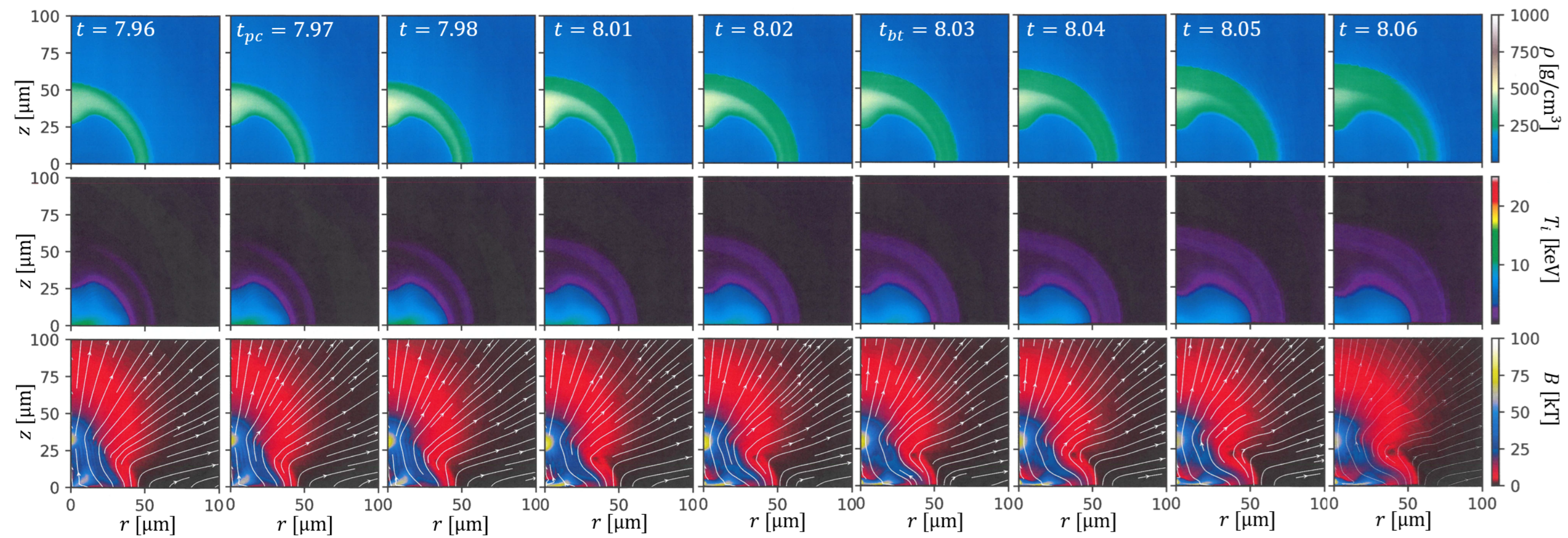}
        \caption{$B=30$ T, peak compression at 7.97 ns and bangtime at 8.03 ns.}
        \label{fig:sub2}
    \end{subfigure}
    \begin{subfigure}{\linewidth}
        \centering
        \includegraphics[width=0.99\textwidth]{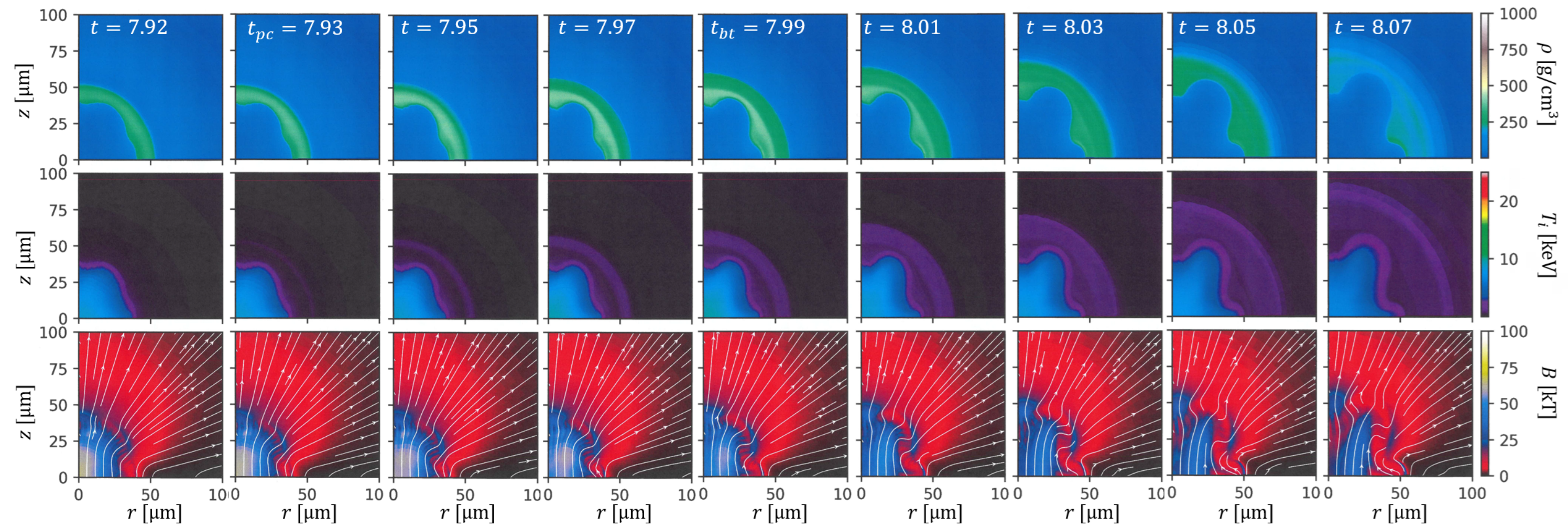}
        \caption{Symmetrized, $B=50$ T, peak compression at 7.93 ns and bangtime at 7.99 ns.}
        \label{fig:sub2}
    \end{subfigure}
\caption{\label{fig:tscan_180128} Snapshots in time of the density $\rho$, ion temperature \Ti, and magnetic field $|B|$ profiles for N180128 when (a) unmagnetized, (b) magnetized at 30 T, and (c) magnetized at 50 T while symmetrized. $t_{pc}$ is the time of peak compression and $t_{bt}$ is the time of bangtime.}
\end{figure*}


\subsection{N210808 - Lawson criterion shot}

A significant advance in the LLNL ICF program occurred in 2021, when NIF shot N210808 exceeded a fusion yield of 1 MJ for the first time.\cite{abu1}  This shot was part of the HYBRID-E campaign,\cite{hurricane} which built upon BigFoot and other campaigns with HDC ablators.  Its goal was to increase yield by coupling more energy to a larger capsule and pushing hohlraum performance further towards its limit, e.g., a smaller LEH.

Performing a magnetic field strength scan, we see that the yield increases significantly, by a factor >6$\times$ for $B=65$ T, as seen in Fig. \ref{fig:bsa} (b). However, the burn-averaged \Ti does not increase as dramatically, with just a 50\% increase in line with what was previously seen for N180128. 
N210808 is distinct in that the symmetrized, magnetized yield is lower than the unsymmetrized, magnetized yield for lower applied field strengths (but still higher than the unmagnetized yield) except near $B = 65$ T, where the enhancement goes up from 6$\times$ to 8$\times$. This may suggest that the asymmetric nature of the implosion, such as the polar caps, was actually more conducive to yield enhancement. However, the compression of a symmetrized implosion is still the superior result when considering peak magnetic enhancement. 

To get a better understanding of the implosion we can again look at lineouts at bangtime ($\sim$9.26 ns), as seen in Fig. \ref{fig:los} (b), comparing $B=0$ T and $B=65$ T. Note the lineouts in Fig. \ref{fig:los} (b) are normalized differently than in Fig. \ref{fig:los} (a). This time the profiles look notably different. The implosion is actually less converged when magnetized on the equator but on the poles the density profile has been completely blown through in a highly asymmetric fashion. One can see that not only is the temperature almost 3$\times$ higher at this time but that there is significant temperature disequilibration between the ions and electrons. On the pole an expansion shock has formed, which is characterized by high ion temperatures relative to that of the electrons. This is due to hydrodynamic viscosity effects independent of magnetization,\cite{velikovich} where shocks preferentially heat ions. In a robustly igniting target, this occurs not only during the implosion but also during the detonation phase, although there are additional second-order effects of magnetization on shock dynamics as well. Again, magnetic field strengths approach 50 kT particularly near the gas-ice interface, where internal ablation of the fuel layer piles up the B-field inward.

\begin{figure*}
    \begin{subfigure}{\textwidth}
        \centering
        \includegraphics[width=0.99\textwidth]{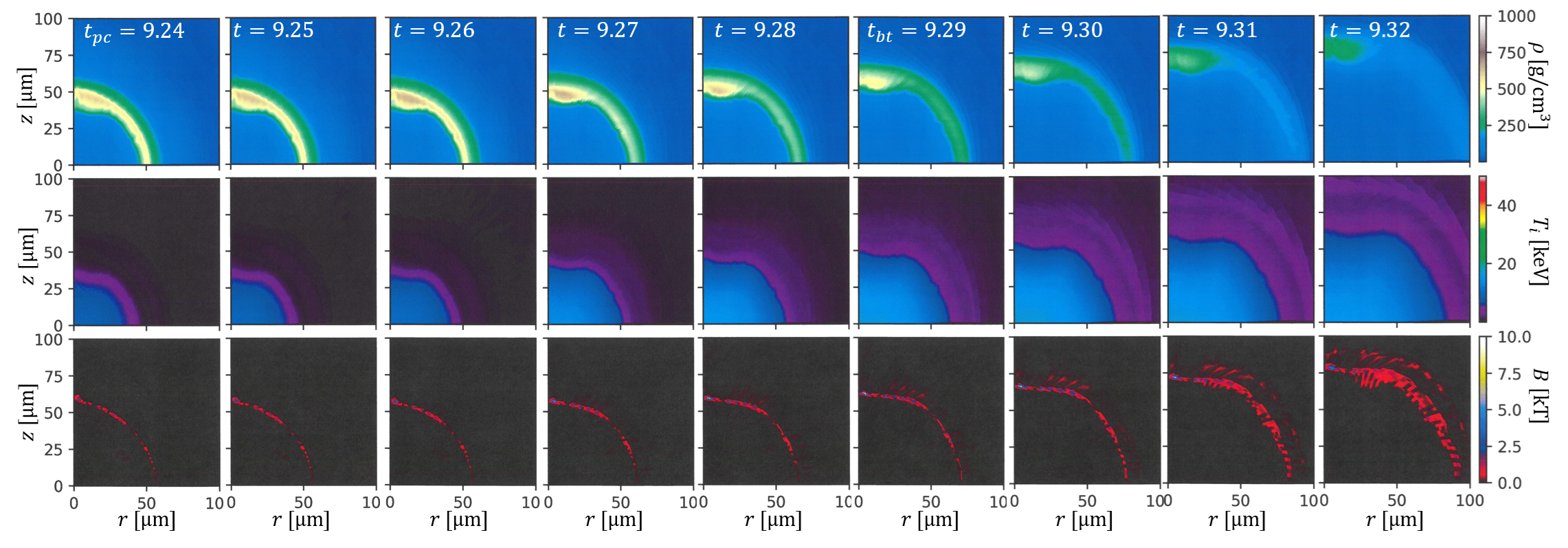}
        \caption{$B=0$, peak compression at 9.24 ns and bangtime at 9.29 ns.}
        \label{fig:sub1}
    \end{subfigure}
    \begin{subfigure}{\linewidth}
        \centering
        \includegraphics[width=0.99\textwidth]{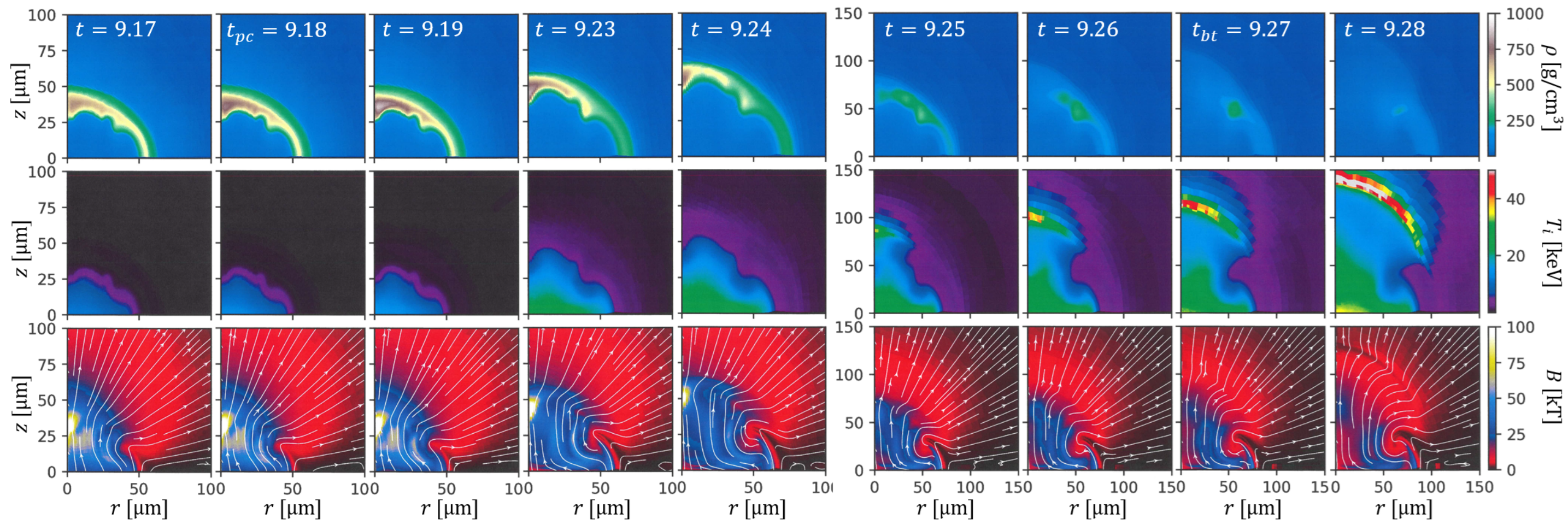}
        \caption{$B=65$ T, peak compression at 9.18 ns and bangtime at 9.27 ns.}
        \label{fig:sub2}
    \end{subfigure}
    \begin{subfigure}{\linewidth}
        \centering
        \includegraphics[width=0.99\textwidth]{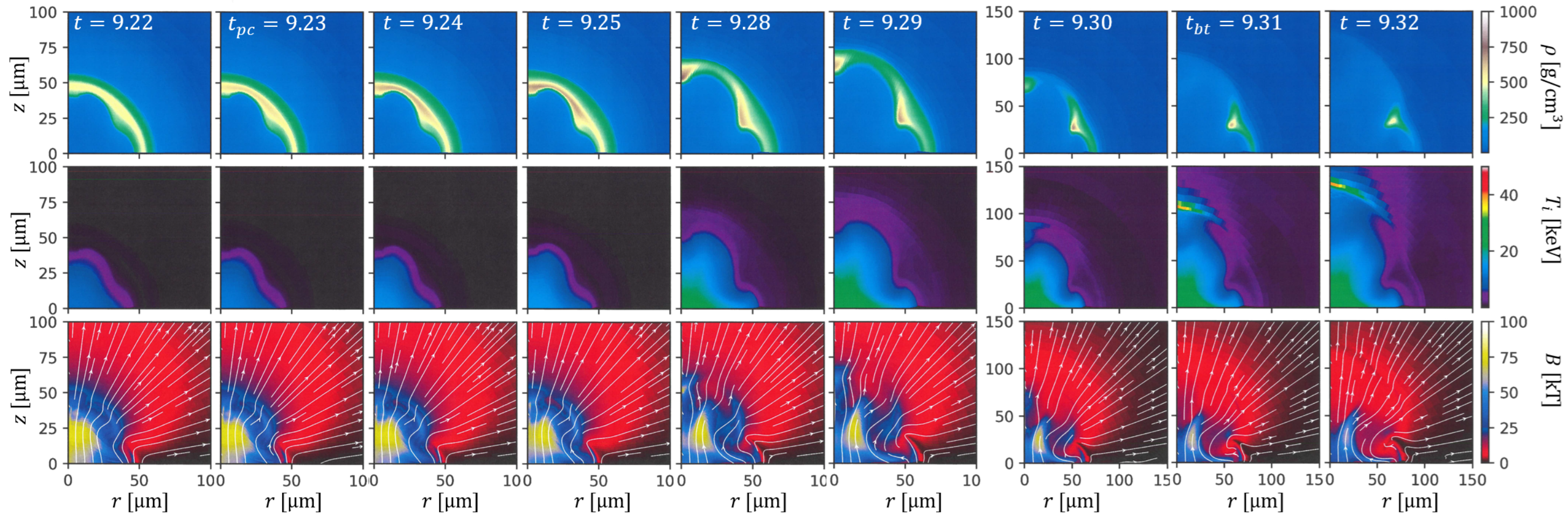}
        \caption{Symmetrized, $B=65$ T, peak compression at 9.23 ns and bangtime at 9.31 ns.}
        \label{fig:sub2}
    \end{subfigure}
\caption{\label{fig:tscan_210808} Snapshots in time of the density $\rho$, ion temperature \Ti, and magnetic field $|B|$ profiles for N210808 when (a) unmagnetized, (b) magnetized at 65 T, and (c) magnetized at 65 T while symmetrized.}
\end{figure*}

To highlight the evolution of the magnetized case example we have plotted the density, temperature, and magnetic field profiles at sequential times between peak compression and bangtime, in Fig. \ref{fig:tscan_210808} (a) for $B = 0$ T and Fig. \ref{fig:tscan_210808} (b) for $B = 65$ T where magnetic enhancement is greatest. We can see the clear formation of polar caps in both cases, typical of current NIF designs, but in the magnetized case one can clearly see the polar cap blowing out early relative to bangtime at 9.26 ns, which seems to be a common occurrence in high yield, magnetized designs. This is reasonable given that there are greater heat conduction and alpha propagation on the poles versus the equator. The unusual polar burn-through of hotspot profile at bangtime can be readily explained by examining the implosion several tens of picoseconds before bangtime, i.e., at 9.23 ns when unmagnetized and 9.22 ns when magnetized. In Fig. \ref{fig:tscan_210808} (a) and (b) we can see 2D images of the density for $B=0$ T and $B=65$ T at relative times and we can observe that the capsule shell is still intact for the magnetized case. In both cases we see the denser polar caps that are typical of HYBRID-E designs born out of the $p_2$ and $p_4$ swings inherent to laser-driven hohlraums. These swings are due to the LEH closure affecting laser propagation and the gold bubble driven by the outer laser cones absorbing energy from the inner cones. Symmetrized results can be seen in Fig. \ref{fig:tscan_210808} (c), which show that magnetic fields result in a $p_4,p_6$-like structure in the density and temperature profiles.

\begin{figure}
\includegraphics[width=\columnwidth]{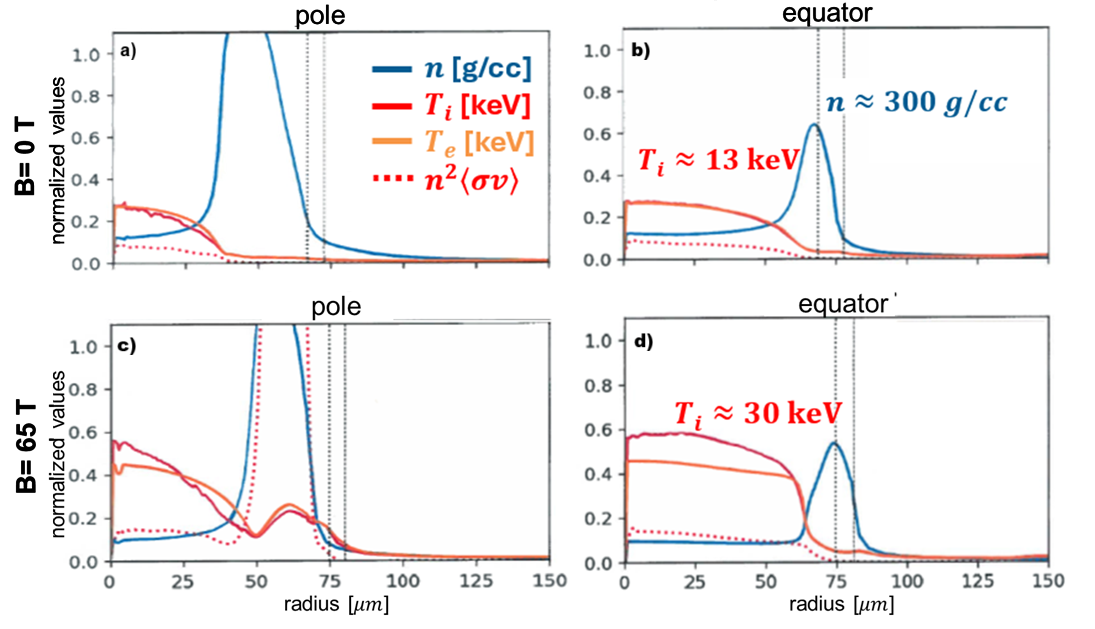}
\caption{\label{fig:lo_210808_comp} Lineouts of N210808 approximately 50 ps before bangtime when (a), (b) unmagnetized ($t=9.23$ ns) and when (c), (d) magnetized at B=65 T applied field ($t = 9.22$ ns). In addition, we have taken lineouts along the pole in (a), (c) and the equator (b), (d). Depicted are the density (blue), \Ti (red), \Te (orange), and nuclear reaction rate (dashed red). The vertical dashed lines define the region of mix between DT and HDC ablator. Quantities in this plot have been normalized as follows: $\rho$ - 1000 g/cm$^{3}$, \Ti - 50 keV, \Te - 50 keV, and $n^2 \langle \sigma v \rangle$ - $10^{-11}$ reactions cm$^{-3}$ s$^{-1}$.}
\end{figure}

Again, we can take lineouts along the pole and equator to more clearly identify what is occurring approximately 50 ps before bangtime. As seen in Fig. \ref{fig:lo_210808_comp}, we can actually find relatively well-behaved implosions for both the magnetized and unmagnetized cases. The density lineouts suggest very similar implosions with significant polar caps. However, the temperature in the magnetized case is not only >2$\times$ in value with respect to the unmagnetized case, but there is also significant heat transport into the poles when magnetized. The high temperatures conspire with the high densities along the poles to effectively ignite the polar caps to an even higher degree than the baseline hotspot itself.  

We can focus on bangtime for the unmagnetized and magnetized cases to compare the various physics parameters of interest in 2D as well, as presented in Fig. \ref{fig:tscan_180128}. We can see when unmagnetized the polar cap is still intact unlike when magnetized at 65 T. \Ti is much hotter in the hotspot when magnetized but also we can see a high temperature band on the pole, which corresponds to a supersonic shock as the implosion detonates on the pole due to the polar cap igniting from enhanced heat flow. In the magnetic field profiles, we can see the formation of self-generated fields on the order of a few kT, particularly when unmagnetized in Fig. \ref{fig:tscan_180128} (a). Self-generated fields form due to imperfections and asymmetries in the implosion via the Biermann battery effect, $\partial B/\partial t = \nabla T_e \times \nabla n_e/n_e e$, whether physical or numerical, and have a tendency to smear out the density profile at stagnation, but not enough to strongly perturb the implosion, with or without applied magnetic fields.


\subsection{N221204 - first breakeven shot}

In 2022, NIF shot N221204 achieved energy breakeven for the first time,\cite{abu2} meaning that more energy was produced via nuclear reactions than delivered by the drive laser to the hohlraum. The most notable change from N210808 was that the laser energy was increased from 1.9 MJ to 2.05 MJ, slightly lengthening the peak laser drive with respect to time. In addition, the thickness of the capsule ablator was slightly increased, beam pointing and $\Delta \lambda$ were retuned to control implosion shape. As noted before, our prediction of the yield was somewhat high, 4.6 MJ versus the experimental 3.2 MJ, but we kept this result since recent experiments based on the N221204 design have since exceeded the original yield studied here.

\begin{figure}
\includegraphics[width=\columnwidth]{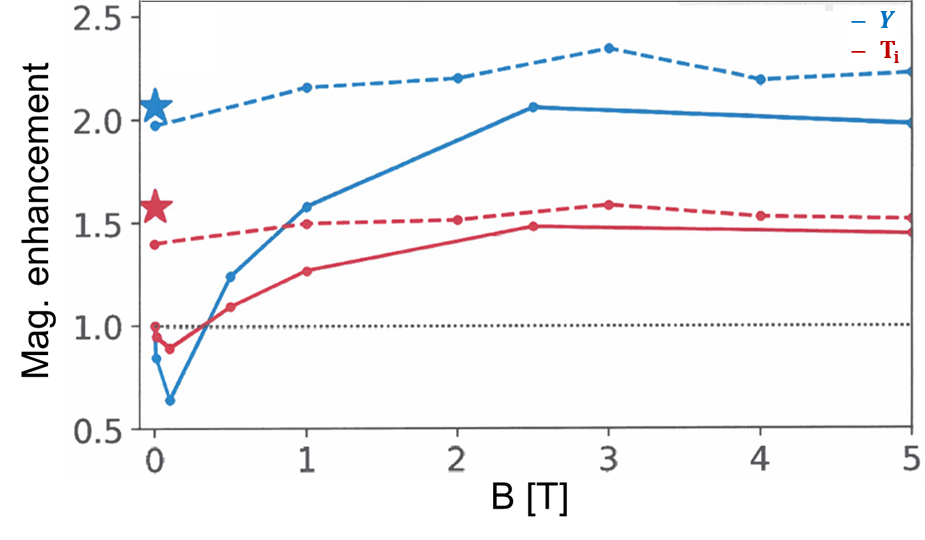}
\caption{\label{fig:221204_low_B} Effect of magnetization at low magnetic field values for N221204. Yield (blue) and \Ti (red) are plotted, where solid lines depict fully integrated simulations, dashed for symmetrized results. The stars depict the symmetrized, unmagnetized result without mix.}
\end{figure}

Applying a magnetic field to N221204 results in yield enhancement of a little more than 2$\times$, not to the same degree as for N210808, which can be seen in Fig. \ref{fig:bsa} (c). In this case, the peak magnetic enhancement occurs at $B=60$ T, analogous to N210808. 
However, we can see a rapid increase in yield enhancement for only a few T of magnetic field strength. There is also, in fact, a noticeable drop at $B=0.1$ T before yield enhancement rises again and levels off around $B=5$ T, observable in Fig. \ref{fig:221204_low_B}. One explanation for the rapid magnetization is that the plasma is already quite hot, which means that it can magnetize more readily than a cooler plasma, although this was not the case for N180128. 

\begin{figure*}
    \begin{subfigure}{\textwidth}
        \centering
        \includegraphics[width=0.99\textwidth]{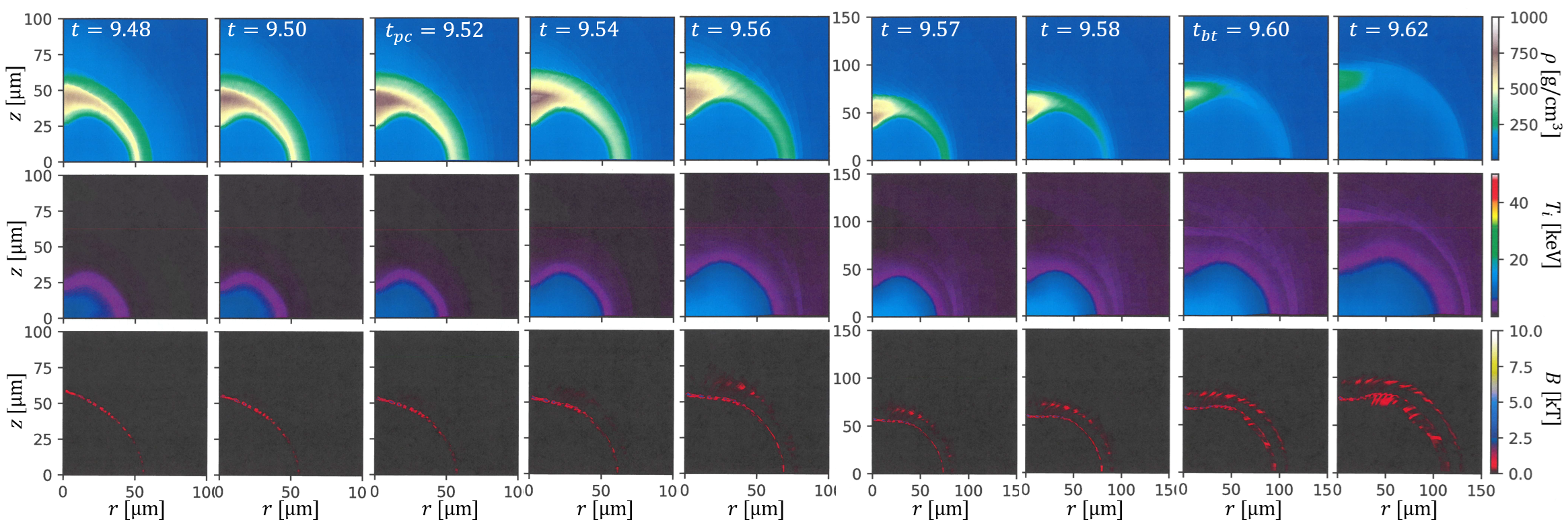}
        \caption{$B=0$, peak compression at 9.52 ns and bangtime at 9.60 ns.}
        \label{fig:sub1}
    \end{subfigure}
    \begin{subfigure}{\linewidth}
        \centering
        \includegraphics[width=0.99\textwidth]{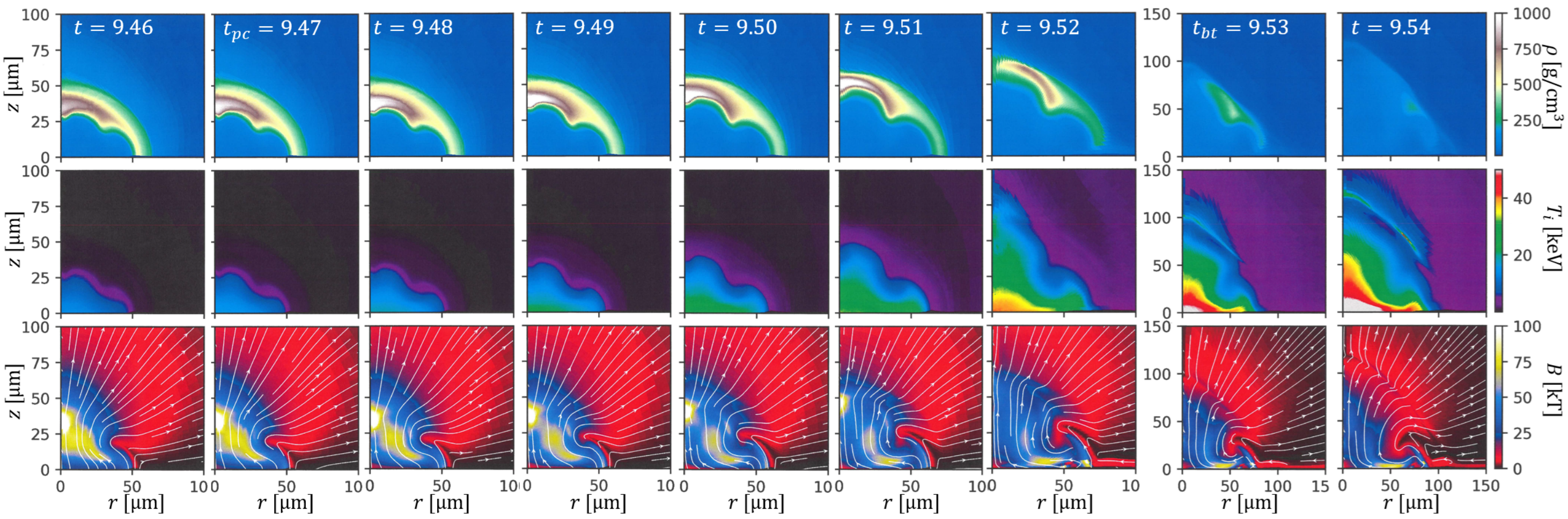}
        \caption{$B=60$ T, peak compression at 9.47 ns and bangtime at 9.53 ns.}
        \label{fig:sub2}
    \end{subfigure}
    \begin{subfigure}{\linewidth}
        \centering
        \includegraphics[width=0.99\textwidth]{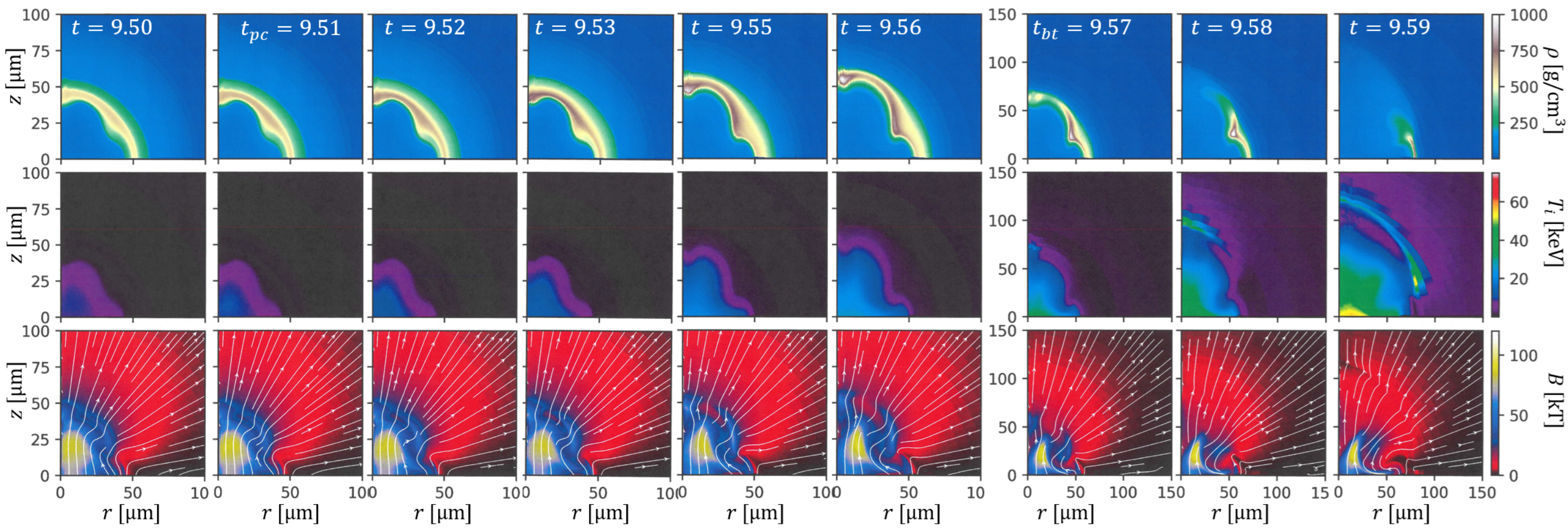}
        \caption{Symmetrized, $B=70$ T, peak compression at 9.51 ns and bangtime at 9.57 ns.}
        \label{fig:sub2}
    \end{subfigure}
\caption{\label{fig:tscan_n221204} Snapshots in time of the density $\rho$, ion temperature \Ti, and magnetic field $|B|$ profiles for N221204 when (a) unmagnetized, (b) magnetized at 60 T, and (c) magnetized at 70 T while symmetrized.}
\end{figure*}

Time series of $\rho$, \Ti, and $|B|$ are shown in Figs. \ref{fig:tscan_n221204} (a)-(c) for a fully-integrated simulation of N221204 unmagnetized, magnetized at $B = 60$ T, and symmetrized at $B = 70$ T, respectively. 
In Fig. \ref{fig:los} (c) are shown lineouts at bangtime, where we have an approximately 20 keV plasma in the hotspot. In this case we have a more symmetric implosion and one in which the implosion dynamics do not differ that much when magnetized. The nuclear reaction rate is already relatively high when unmagnetized and effectively doubles when magnetized, directly correlating to the approximate 2$\times$ increase in yield. Notably, the magnetic field strength starts to approach 100 kT.

With symmetrization N221204 doubles in yield even without an applied magnetic field, suggesting that shape improvements or reduction in residual kinetic energy could significantly enhance the yield. Mix, however, was a minor factor, given that the no-mix, symmetrized result does not visibly differ. However, the peak enhancement still comes with magnetization, reaching $3\times$ in yield enhancement at 70 T with respect to the unmagnetized, unsymmetrized data point but only a 50\% increase relative to the unmagnetized, symmetrized zero point. This suggests that laser drive symmetry can be improved even further without an applied magnetic field but that it is still beneficial to apply one. However, achieving such a perfect implosion is currently impossible, as the $p_2$ and $p_4$ swings in the x-ray drive will result in some asymmetry, i.e., polar caps. It is possible that magnetic fields can synergize with such asymmetries however.

\begin{figure}
\includegraphics[width=\columnwidth]{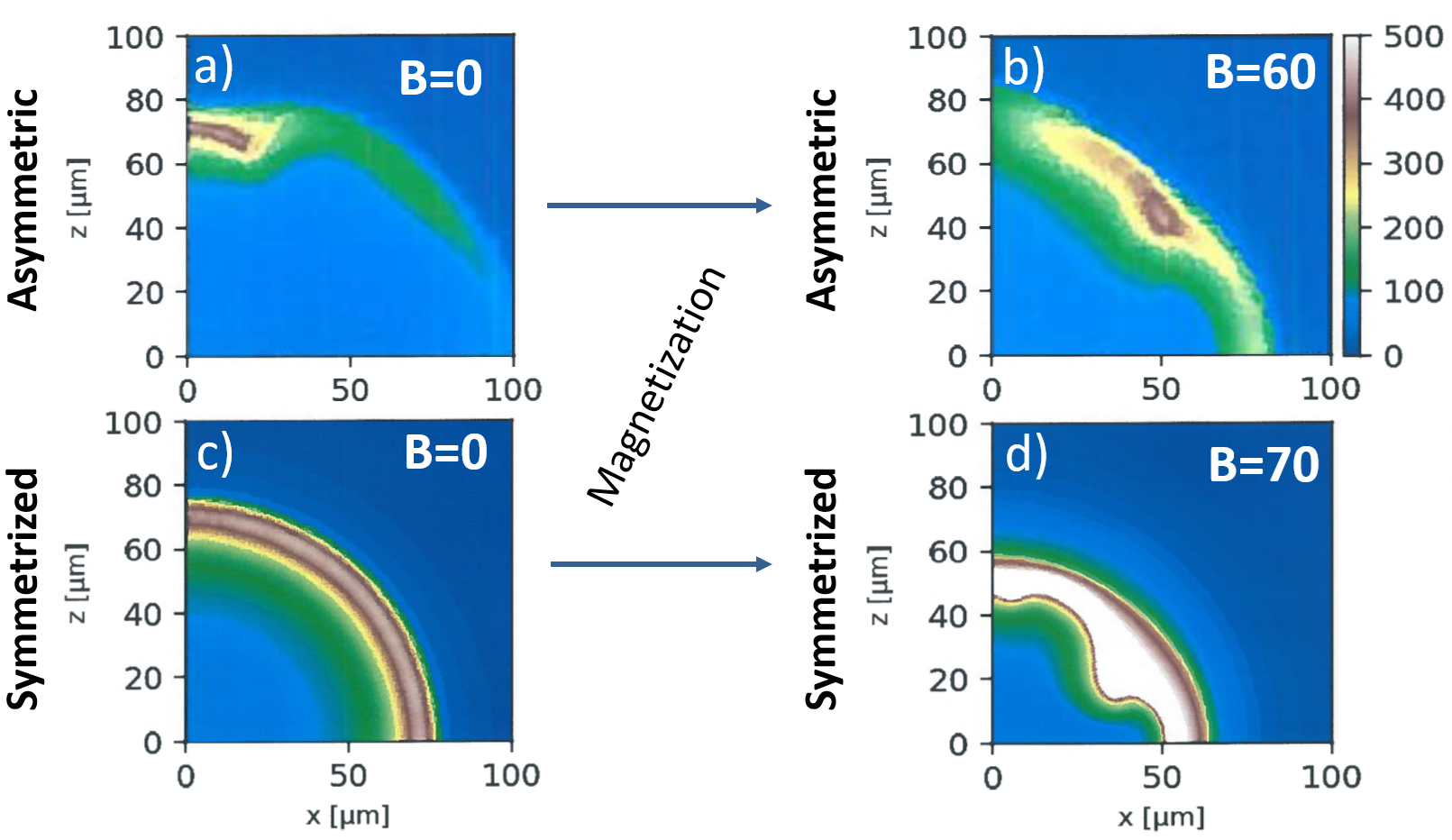}
\caption{\label{fig:compsym22} A comparison of N221204 when (a) unmagnetized and unsymmetrized, (b) magnetized and unsymmetrized, (c) unmagnetized and symmetrized, and (d) magnetized and symmetrized.}
\end{figure}

For N221204 we can compare the profiles of the implosion with and without symmetrization at their respective peak yield enhancement points, as visible in Fig. \ref{fig:compsym22}. When symmetrized the implosion converges somewhat more and thereby reaches higher densities. Noticeably, the applied magnetic field still induces asymmetries to arise at stagnation. However, the polar cap does not get burned through when symmetrized at $B=70$ T unlike when unsymmetrized at $B=60$ T. 

Why such a difference occurs between N210808 and N221204 is still under investigation. N221204 is a more robust and symmetric implosion than N210808, so the effect of magnetization is less. Likewise, the level of mix is lower for N221204 versus N210808, but arbitrarily switching the mix values between the two does not significantly change the results for yield or \Ti. Likewise, the rapid rise in yield for N221204 cannot just be attributed to the fact that hotter plasmas magnetize more readily, as N180128 demonstrates similar rapid magnetization unlike N210808.


\section{Pushered Single Shell (PSS)} \label{pss}

PSS is an alternative approach to HYBRID-E whereby high-Z materials are used in the inner capsule layer in order to enhance confinement and radiation trapping and lower ignition requirements. In PSS we have a relatively dense, thick shell that has a density gradient throughout its radius as previously depicted in Fig. \ref{fig:targets}. The shape of the Mo concentration $f_\text{Mo}$ is given by,\cite{maclaren3}
$$f_\text{Mo}(R) = f_\text{Mo,peak} \left( \frac{\tanh[a(1-R)]-\tanh(ab)} {\tanh[a(1-b)]-\tanh(ab)} \right),$$
where $R=\frac{r-r_i-r_1}{r_2}$ is the normalized radius, $r_i=970 \ \mu$m, $r_1=1 \ \mu$m, $r_2=30 \ \mu$m are the inner ablator radius, dopant inner plateau thickness from $r_i$, and the end of the graded layer from $r_i$, respectively, and the coefficients are $a=2.2$ and $b=-0.1$. The PSS approach mitigates $PdV$ loss through its unique ablator design, providing more remaining mass to tamp the hot spot during ignition and burn. This design is particularly interesting because it achieves large $\rho R_f$ values at stagnation, which means more fuel for alphas to interact with. However, PSS generally only achieves low hotspot temperatures. The idea here is that we will use this design to assemble much more fuel but also with a slower implosion velocity than typically found in a HYBRID-E design, meaning longer confinement time. We will compensate for the baseline lower temperature by using an applied magnetic field to raise it. 

One notable risk of current PSS designs is the susceptibility to significant mix degradation of the implosion. Bremsstrahlung losses are a major loss in all hotspot ignition designs, but when high-Z components of the inner ablator mix into the hotspot, the high average ionization state of the hot spot mixture can significantly hinder the ability of the hotspot to reach high temperatures. For this reason, we will consider two separate instances of the otherwise identical PSS design, one with moderate levels of mix and another will lower levels of mix. Moderate mix has a mixlength of $\delta h = 0.25 \mu m$ and ablator mix fraction of $f_\text{mix}=0.02\%$ while the low mix instance has $\delta h = 0.2 \mu m$ and $f_\text{mix}=0.015\%$. This a relatively small difference numerically but results in significant differences in performance, both for unmagnetized and magnetized designs. Mix is an outstanding issue in all ICF designs, not just magnetized ones or PSS.

\begin{figure*}
    \begin{subfigure}{\textwidth}
        \centering
        \includegraphics[width=0.99\textwidth]{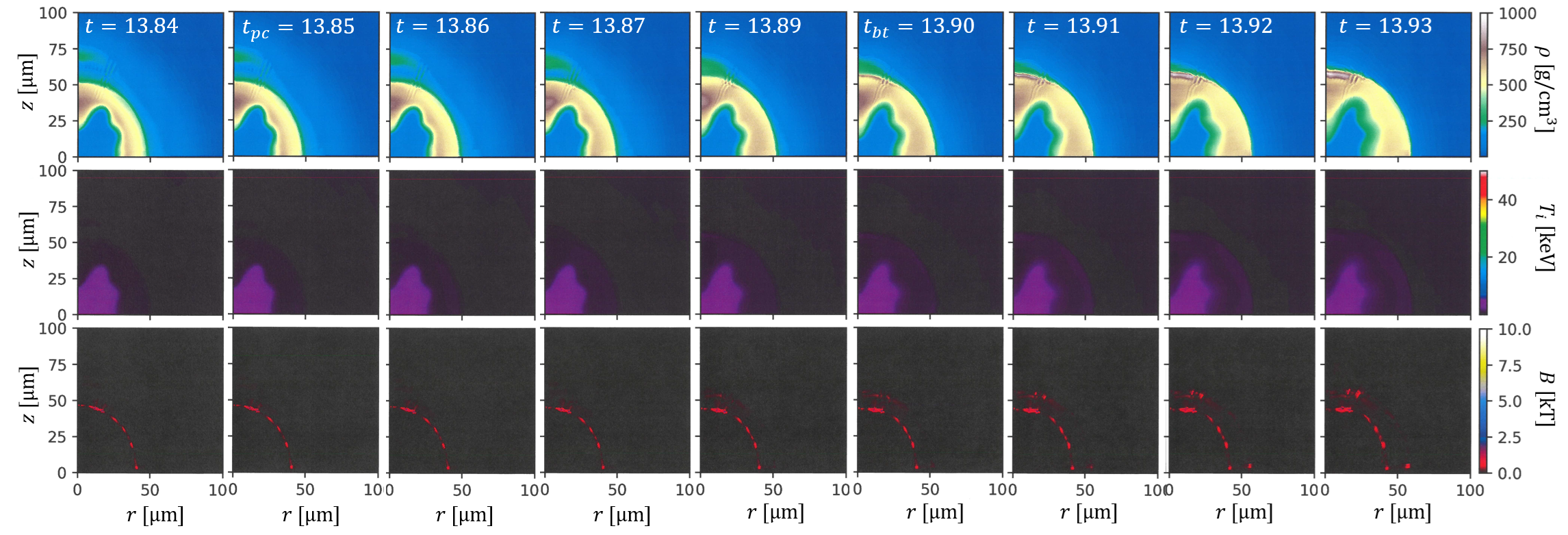}
        \caption{$B=0$, peak compression at 13.85 ns and bangtime at 13.90 ns.}
        \label{fig:sub1}
    \end{subfigure}
    \begin{subfigure}{\linewidth}
        \centering
        \includegraphics[width=0.99\textwidth]{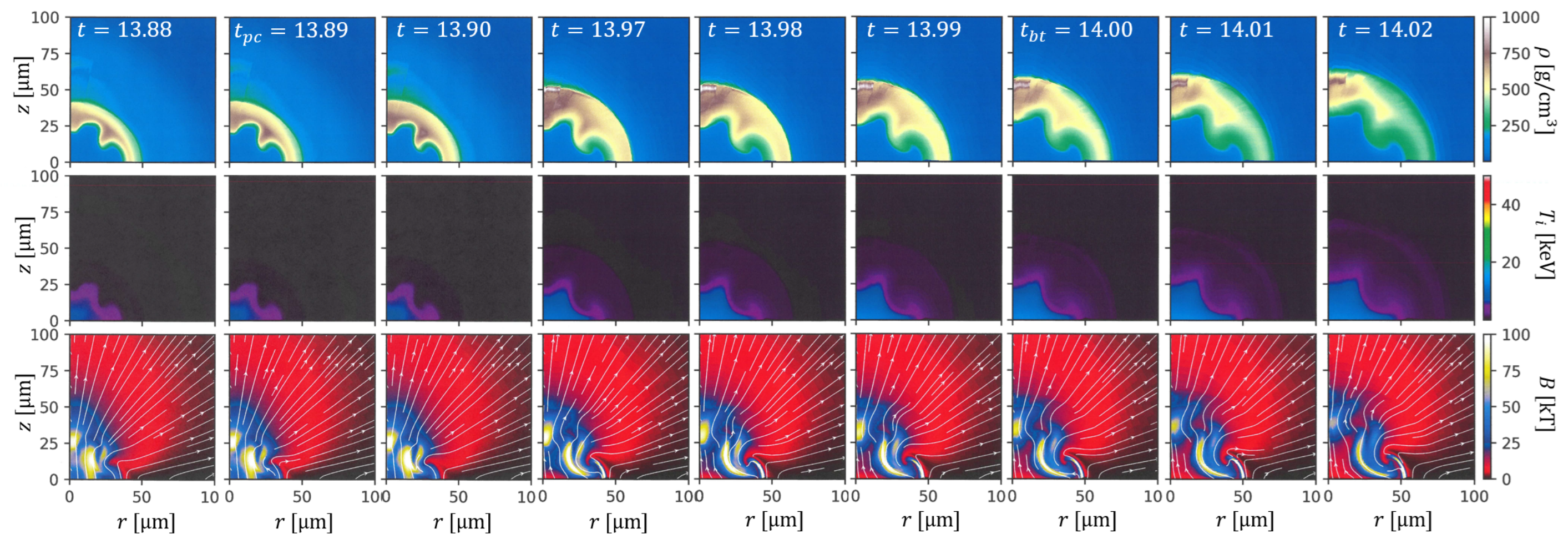}
        \caption{$B=50$ T, peak compression at 13.89 ns and bangtime at 14.00 ns.}
        \label{fig:sub2}
    \end{subfigure}
    \begin{subfigure}{\linewidth}
        \centering
        \includegraphics[width=0.99\textwidth]{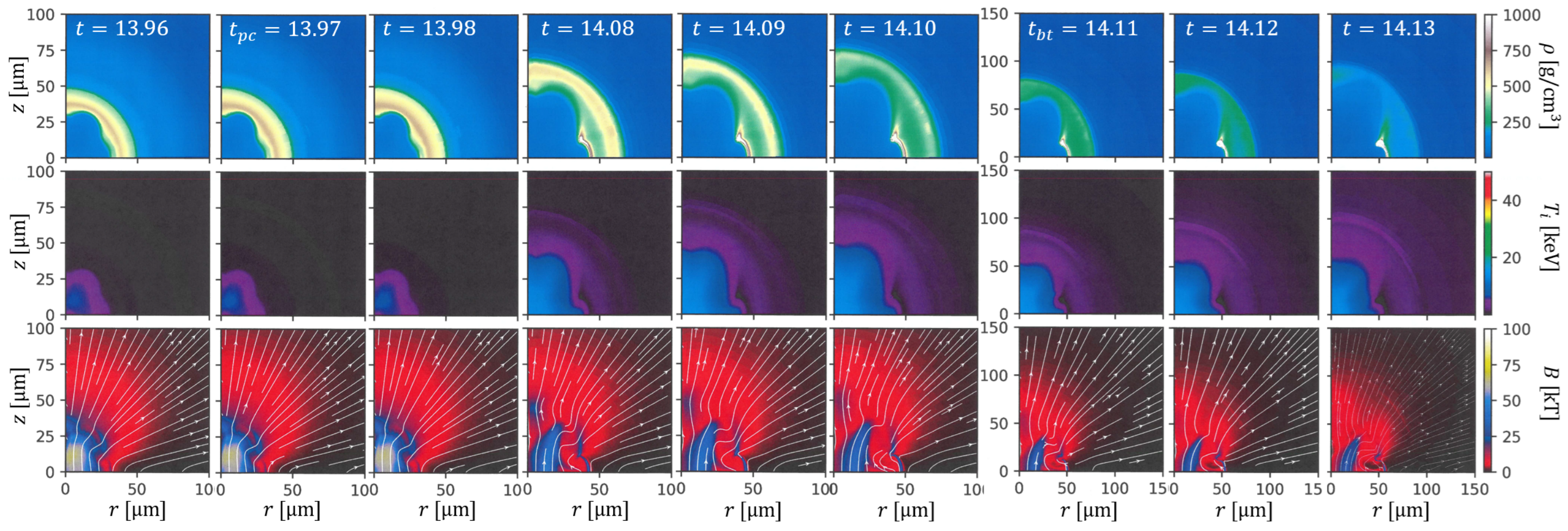}
        \caption{Symmetrized, $B=30$ T, peak compression at 13.97 ns and bangtime at 14.11 ns.}
        \label{fig:sub2}
    \end{subfigure}
\caption{\label{fig:tscan_pss_med} Snapshots in time of the density $\rho$, ion temperature \Ti, and magnetic field $|B|$ profiles for PSS with moderate mix parameters  ($\delta h =0.25\ \mu$m, $f_\text{mix}=0.02 \%$) when a) unmagnetized, b) magnetized at 50 T, and c) magnetized at 30 T while symmetrized.}
\end{figure*}

\begin{figure*}
    \begin{subfigure}{\textwidth}
        \centering
        \includegraphics[width=0.99\textwidth]{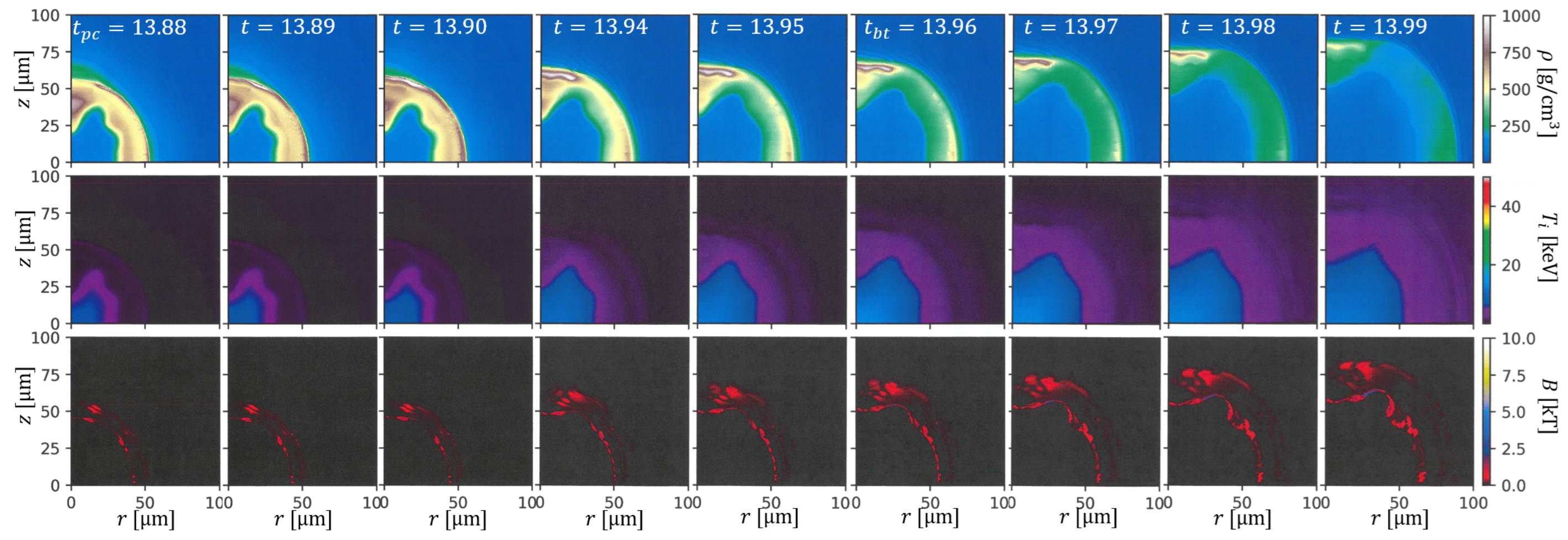}
        \caption{$B=0$, peak compression at 13.88 ns and bangtime at 13.96 ns.}
        \label{fig:sub1}
    \end{subfigure}
    \begin{subfigure}{\linewidth}
        \centering
        \includegraphics[width=0.99\textwidth]{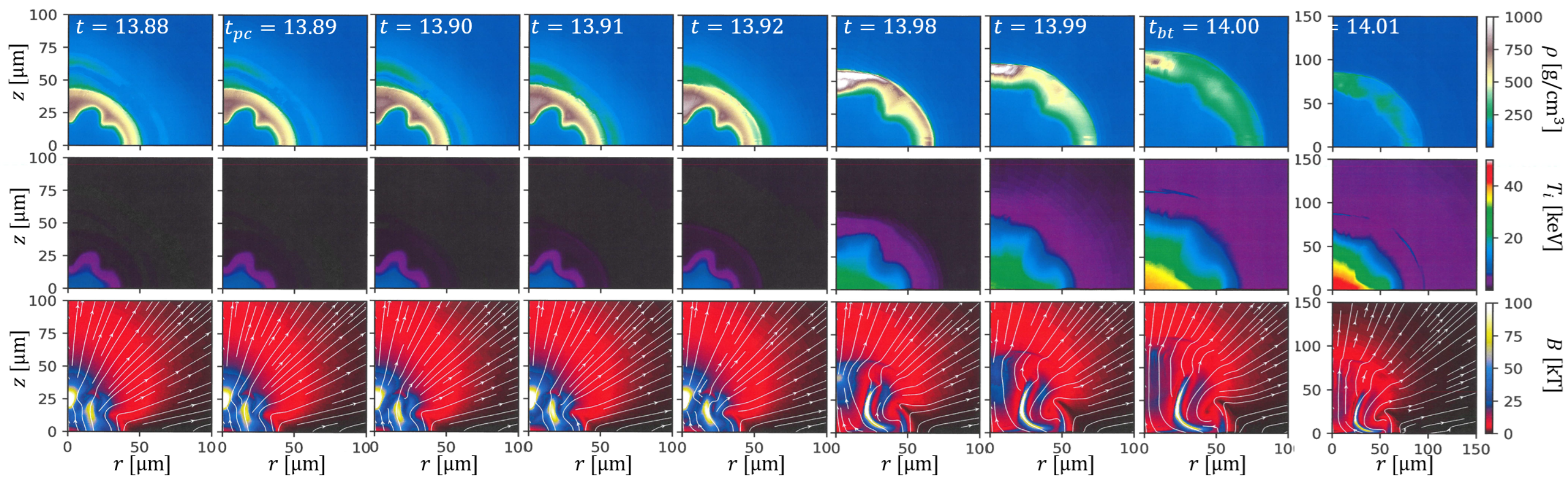}
        \caption{$B=30$ T, peak compression at 13.89 ns and bangtime at 14.00 ns.}
        \label{fig:sub2}
    \end{subfigure}
    \begin{subfigure}{\linewidth}
        \centering
        \includegraphics[width=0.99\textwidth]{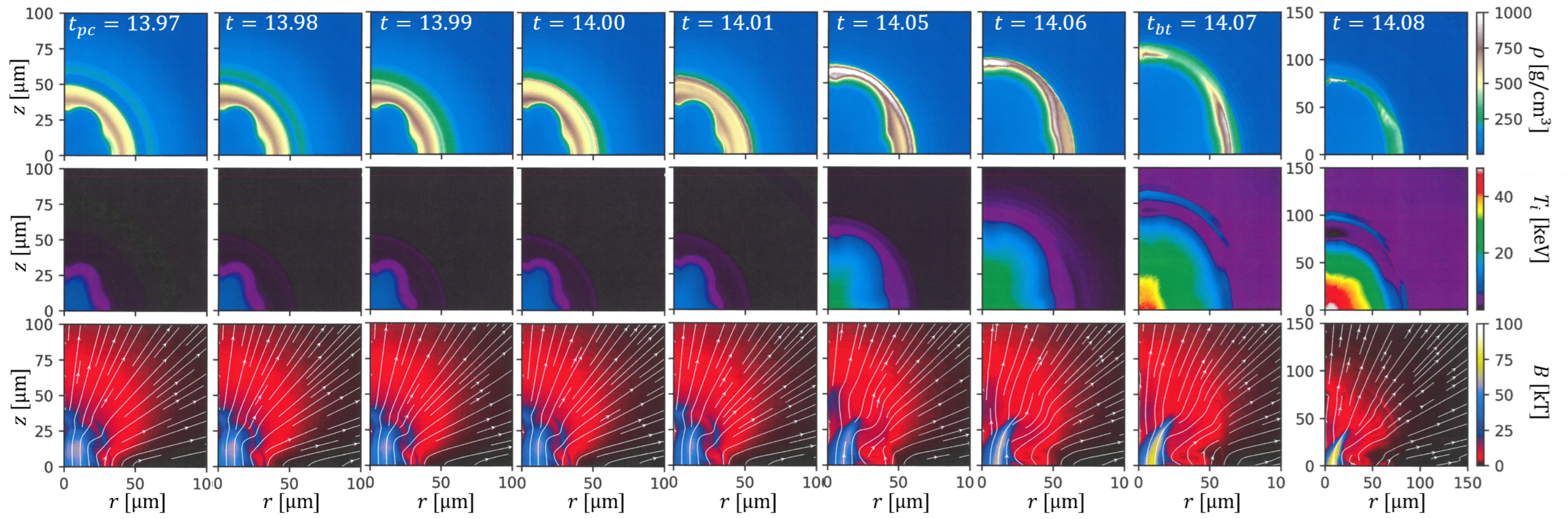}
        \caption{Symmetrized, $B=30$ T, peak compression at 13.97 ns and bangtime at 14.07 ns.}
        \label{fig:sub2}
    \end{subfigure}
\caption{\label{fig:tscan_pss_low} Snapshots in time of the density $\rho$, ion temperature \Ti, and magnetic field $|B|$ profiles for PSS with low mix parameters ($\delta h =0.2\ \mu$m, $f_\text{mix}=0.015\%$) when (a) unmagnetized, (b) magnetized at 30 T, and (c) magnetized at 30 T while symmetrized.}
\end{figure*}

We first perform a magnetic field scan for PSS with a 1.9 MJ laser drive and medium mix, which is closer to what has been observed in radiochemistry experiments thus far. We see a pronounced $\sim$4$\times$ peak in yield enhancement at 50 T as seen in Fig. \ref{fig:bsa} when unsymmetrized and $\sim$12.5$\times$ at 30 T when symmetrized, going from 63 kJ to 253 kJ and 788 kJ, respectively. For both moderate and low mix, 
a visualization of the time evolution can be seen in Figs. \ref{fig:tscan_pss_med} and \ref{fig:tscan_pss_low}. When lowering mix model parameters by $\sim$20 \%, we see significant enhancement in yield both when magnetized and unmagnetized. For an asymmetric and symmetric drives we see a peak magnetic enhancement at 30 T, going from 0.97 MJ to 6.3 MJ (6.5$\times$) and to 9.7 MJ (10$\times$) for asymmetric and symmetric drives, respectively. The significant enhancement due to symmetrization suggests that a PSS design is a promising candidate for further optimization, although in addition to mix it was also found to be more sensitive to shape asymmetries than HYBRID-E designs. However, these are baseline design questions that do not preclude future magnetization of them, and there exists the possibility that $p_4$ asymmetries caused by magnetization may balance out those caused by hydrodynamic asymmetries.

\begin{figure}
\includegraphics[width=\columnwidth]{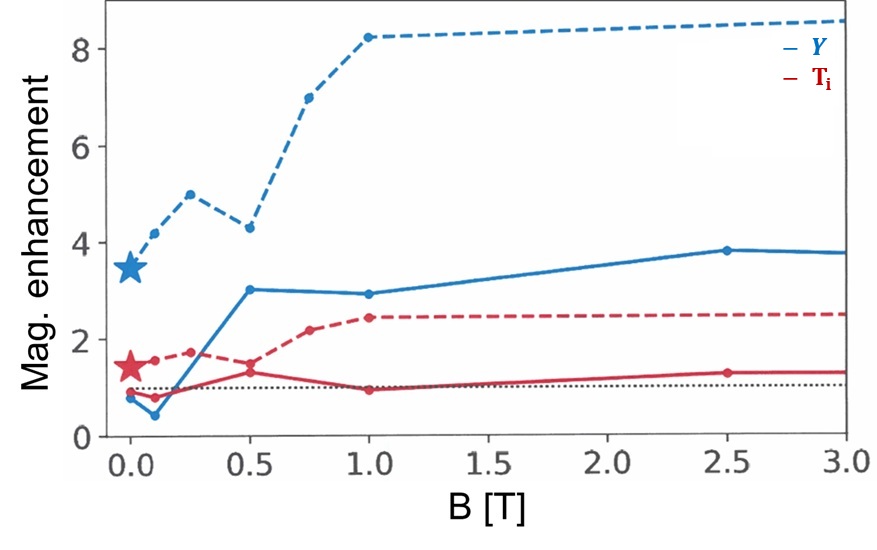}
\caption{\label{fig:pss_low_B} Effect of magnetization at low magnetic field values for the low-mix PSS design.  Yield is plotted in blue and \Ti in red. Solid lines depict fully integrated simulations, dashed for symmetrized results. The stars depict the symmetrized result without mix.}
\end{figure}

If we look at lineouts of the implosion at bangtime, Fig. \ref{fig:los} (e) in this case we can see that we assemble on average higher densities and that we are able to more than double the temperature within the hotspot. Magnetization addresses one of the primary issues of the PSS design, lower hot spot temperatures, while also leveraging its advantageous, higher $\rho R$, assembling more fuel at stagnation. The potential benefit of magnetization alleviating mix levels may further benefit the magnetized-PSS concept, given its sensitivity to mix. Likewise, as was the case for N221204, we see significant enhancement even with the application of very low strength magnetic fields. This can be seen in Fig. \ref{fig:pss_low_B}, where an only a 1 T external field results in >2$\times$ increase in yields for both the symmetrized and asymmetric drive cases for the low-mix PSS design.

\section{Enhanced Yield Capability} \label{eyc}

Ongoing efforts in the ICF community have been looking at enhancing the energy of the NIF laser from 2.2 MJ to 2.6 MJ by the 2030s, with potential for going up to 3 MJ thereafter. In this study we consider a 3 MJ EYC design as presented in Ref. \onlinecite{maclaren1}. In this design, the hohlraum is wider and longer given the longer laser pulse and has a larger LEH than current high-performing designs. This is necessary to control symmetry for the larger capsule as well as prevent Au LEH plasma blow-off from interfering with the 44$^\circ$ and 50$^\circ$ laser beams. There is a higher picket at the beginning of the laser pulse which complicates symmetry. The injection of high power at the beginning of the laser pulse is needed for the first shock to completely melt the HDC ablator. However, this causes faster growth of the gold bubble on the hohlraum wall, causing more of the inner cone power to be absorbed closer to the outer beam spots, pushing the capsule symmetry more towards an oblate implosion.\cite{callahan1} This is counteracted by applying greater wavelength separation between the inner (23$^\circ$,30$^\circ$) and outer (44$^\circ$,50$^\circ$) beams, going from $\Delta\lambda = 2.75$ \r{A} to $4$ \r{A}. A challenge with the 3 MJ HDC design used in this study is that one must overcome radiative losses caused by mixing of the carbon ablator into the hotspot. This requires a higher implosion velocity than necessary for previous designs. Increased velocity provides the necessary hotspot energy and density to overcome bremsstrahlung losses, which was achieved by increasing the ablator thickness and reducing the fuel mass. 

Magnetizing the 3 MJ EYC design demonstrates the potential limitations of simple application of external magnetic fields. As seen in Fig. \ref{fig:bsa} (f), the benefit of magnetization for this design is limited if not detrimental. First of all, we were not able to retrieve the full 25 MJ results from Ref. \onlinecite{maclaren1}, achieving at best an initial 13.7 MJ yield at $B=0$ T, which steadily degrades with increased magnetic field strength. The primary difference here is that we used laser power multipliers as opposed to opacity multipliers to be consistent with the rest of the simulations in this study. When symmetrized, which is the ideal result, we are able to achieve 30 MJ. We can see the negligible influence of a magnetic field up to 100 T, with slight degradation in the yield until it rises again at 90 T, before falling precipitously again as we increase field strength up to 200 T. 
The peak in \Ti at 90 T suggests that we might obtain some modest benefit of magnetization with some optimization of the design. A noticeable distinction here is that $T_{rad}$ non-trivially decreases with magnetic field strength as seen in Fig. \ref{fig:magsym}.e). The origin of this is not clear yet and is the subject of ongoing investigation.

\begin{figure*}
    \begin{subfigure}{\textwidth}
        \centering
        \includegraphics[width=0.99\textwidth]{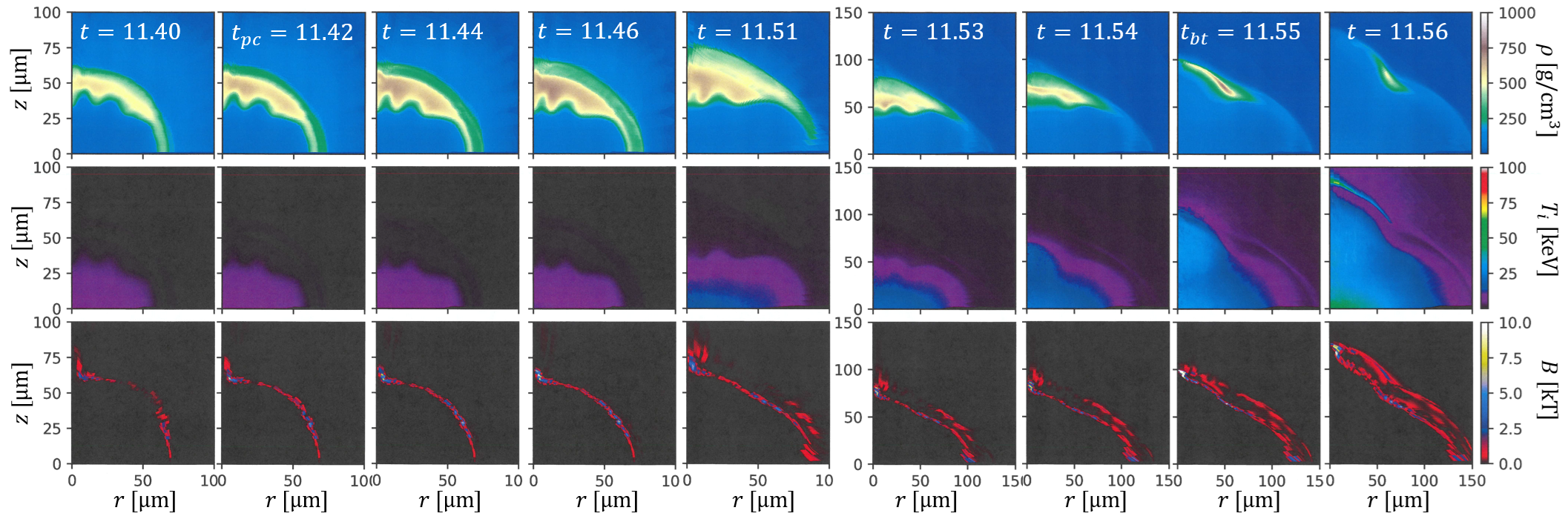}
        \caption{$B=0$, peak compression at 11.42 ns and bangtime at 11.55 ns.}
        \label{fig:sub1}
    \end{subfigure}
    \begin{subfigure}{\linewidth}
        \centering
        \includegraphics[width=0.99\textwidth]{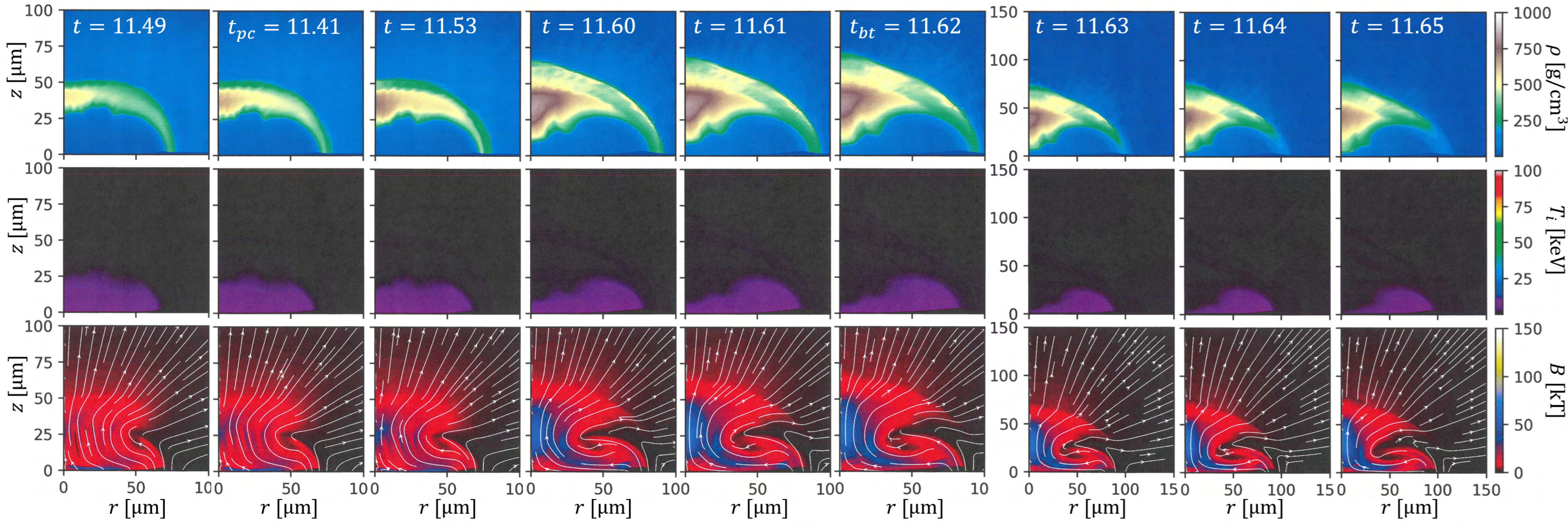}
        \caption{$B=20$ T, peak compression at 11.41 ns and bangtime at 11.62 ns.}
        \label{fig:sub2}
    \end{subfigure}
    \begin{subfigure}{\linewidth}
        \centering
        \includegraphics[width=0.99\textwidth]{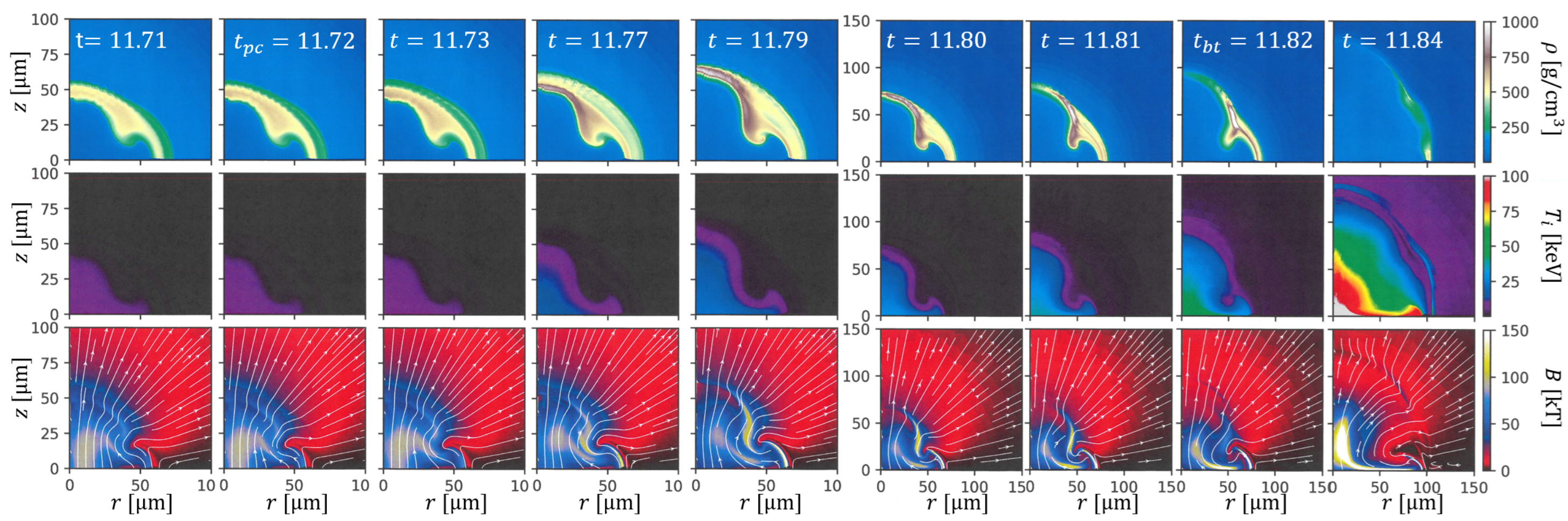}
        \caption{Symmetrized, $B=90$ T, peak compression at 11.72 ns and bangtime at 11.82 ns.}
        \label{fig:sub2}
    \end{subfigure}
\caption{\label{fig:eyc_tscan} Snapshots in time of the density $\rho$, ion temperature \Ti, and magnetic field $|B|$ profiles for EYC when (a) unmagnetized, (b) magnetized at 20 T, and (c) magnetized at 90 T while symmetrized.}
\end{figure*}

Time series depictions of EYC at $B = 0$ T, $B = 20$ T, and then symmetrized at $B = 90$ T can be seen in Fig. \ref{fig:eyc_tscan} (a)-(c), respectively. The unmagnetized case was tuned to have relatively good symmetry and high convergence, as well as clear polar caps to take advantage of magnetized heat transport. Like high-yield shots before, the unmagnetized case achieves high yields that result in an expansion shock forming after bangtime. When we magnetize the implosion, the yield declines. Interestingly, despite the heat confinement from the magnetic field, overall we see lower temperatures even near peak compression at $t_{pc}=11.72$ ns. We can see in the symmetrized case that the magnetic field strongly distorts the shape of the implosion at peak compression and bangtime, again resulting in a $p_4$ structure with a $p_2$ asymmetry causing greater compression on the poles versus the equator. Notably, magnetic field strengths start to approach 200 kT levels and the ion temperature \Ti briefly exceeds 100 keV in the hotspot shortly after bangtime. 


\begin{figure}
\includegraphics[width=\columnwidth]{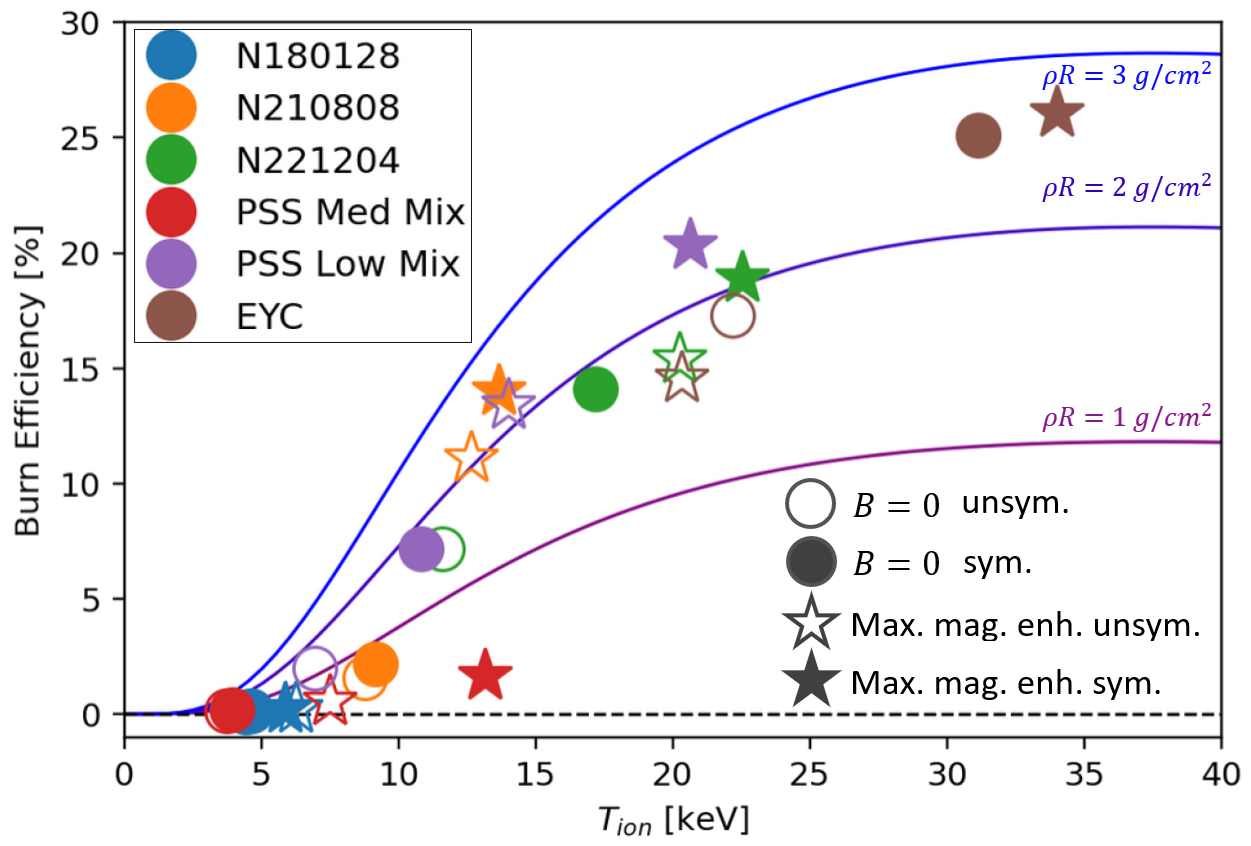}
\caption{\label{fig:burnefficiency} Burn efficiency of the designs considered in this work. The solid lines represent the unmagnetized burn efficiency as a function of \Ti and three different $\rho R$ values (1, 2, and 3 g/cm$^2$) per Eqn. \ref{burneff}. The empty circle is for the baseline, unmagnetized design, solid circle for when symmetrized, the empty star is for the simulation with peak magnetic enhancement, and the solid star for when symmetrized, per Fig. \ref{fig:bsa}.}
\end{figure}

As yield performance increases, we see diminishing returns from applied magnetic fields. The basic understanding is that the hotspot already gets so hot that it does not heat much more from constrained heat conduction and alpha magnetization. There is a natural peak in the cross section $<\sigma\nu>$ of the DT nuclear reaction at about 50 keV, and we see clearly diminishing returns as one exceeds 20 keV. This trend can be seen in Fig. \ref{fig:burnefficiency}, where we plot the burn fraction, i.e., $1-m_{DT,final}/m_{DT,inital}$, versus an analytical model\cite{atzeni} of burn efficiency. 
\begin{equation}
\Phi \approx \rho R_f /(\rho R_f + H_B), 
\label{burneff}
\end{equation}
for the shots in consideration (empty circle), the peak-associated magnetization case (empty star), and the associated symmetrized runs (solid circle and star). Here $m_{DT,i}$ and $\rho R_f$ is the mass and areal density of the imploded fuel, respectively, as it begins to burn and $H_B=20c_sm_p/<\sigma v>$, where $c_s$ is the sound speed. $H_B$ depends only on \Ti and minimizes as $\sim$7 g/cm$^2$ for \Ti$\sim$ 30 keV. In addition, we have plotted out the burn-efficiency curves for constant $\rho R$ values, while for the shots in question we considered the maximum $\rho R$ achieved in the simulation. The estimated maximum burn efficiency is 30\%, which is being approached by N221204, PSS with low mix, and EYC when magnetized. When designs are already achieving very hot temperatures they will not benefit from magnetization, thus the motivation for designs with lower implosion velocity and higher $\rho R$, such as the PSS design.

\section{Conclusion} \label{conclusion}

In this manuscript, we explored the application of external magnetic fields to various high-performing ICF designs. We considered the effect of magnetization on historically high-performing NIF experiments N180128, N210808, and N221204 after rigorously tuning baseline, unmagnetized simulations to experimental data. This used the \lasnex{} Hohlraum Template (LHT), a common model in the Livermore ICF program. In addition, we considered designs of future experiments for Pushered Single Shell (PSS), a promising high $\rho R$ design, as well as a 3 MJ Enhanced Yield Capability (EYC) design to take advantage of near future upgrades to the NIF laser system. 

With the exception of the EYC design, all designs showed some degree of enhancement with applied magnetic fields between 30 to 70 T, with yield enhancement between 2 to 12$\times$. This is primarily due to enhanced hotspot temperatures due to restricted electron heat conduction by the compressed magnetic fields. Alpha magnetization is also an important factor and contributes to the ignition of polar caps when significant at stagnation, e.g., N210808.  Our results on specific NIF designs show that igniting systems can indeed benefit from magnetization, unlike the claim from works like Ref. \onlinecite{jones}. Only for $\geq 100$ T do we start to see a non-retrievable degradation of an igniting design, where shape degradations due to magnetization are too severe for proper assembly of the implosion at stagnation. A unique, unexpected observation was a notable 2-3$\times$ yield enhancement can be achieved by applying magnetic field strengths less than 5 T, even as low as 1 T. This observation has spurred the consideration of low-strength, potentially even permanent-magnetic methods for magnetizing ICF targets. This could be much cheaper than the current pulsed-power method, which is not yet available for cryogenic NIF experiments. 

An ongoing investigation into the fundamentals of magnetized ICF is underway to better understand the nature of these designs. The primary benefit of magnetization comes from a reduction in electron heat conduction perpendicular to magnetic field lines and enhanced alpha energy deposition. However, extended MHD is far more complicated than just these two effects. A follow-up manuscript will be published as Part II of this study, where we perform a cursory survey of extended MHD effects in magnetized ICF implosions. To date, reduced heat conduction and magnetic confinement of energetic ($\sim$MeV) charged fusion products have been demonstrated in NIF experiments. However, the long-considered concept of magnetized mix is still under study and is a potentially promising reason for further pursuit of magnetized ICF. 

Having surveyed well-vetted and established ICF designs and their response to magnetization, ongoing work is being pursued. Given experiences with robustly igniting designs like N221204 or EYC, a wiser application of magnetic fields needs to be considered. One promising avenue is to tune the magnetic field profile in addition to that of a simple axial field, e.g., a magnetic mirror design that might further restrict conduction losses.\cite{walsh3} An additional consideration is to shim the ice layer inside the capsule so that more fuel mass is accrued on the poles, taking advantage of the collimated heat flow and alpha deposition that we see in our simulations. To truly take advantage of magnetized ICF however, it will be necessary to develop new ICF designs that are specifically tailored to the effects of magnetization. A recent study performed scans of 1D capsule-only simulations with reduced conductivity and alpha magnetization coupled to a smaller set of 2D simulations.\cite{walsh4} It was found that a redesign of a non-igniting model of N170601 (measured yield of 42 kJ), increasing the shell thickness and thinning the ice layer, coupled to applied magnetic fields, could result in an 18.5$\times$ enhancement in yield. Such considerations can be extended to fully integrated hohlraum simulations of magnetized ICF implosions. Given the high dimensionality and nonlinearity of the problem, brute-force scans of parameter space would be prohibitively expensive. This suggests an opportunity for the application of numerical optimization, whether using Gaussian processes, adjoint methods, or low-cost gradient-based approaches.


\newpage
\pagebreak

\begin{acknowledgments}
This work was conducted under the auspices of the U.S. Department of Energy by LLNL under contract DE-AC52-07NA27344 and Laboratory Directed Research and Development project 23-ERD-025. This document was prepared as an account of work sponsored by an agency of the United States government. Neither the United States government nor Lawrence Livermore National Security, LLC, nor any of their employees makes any warranty, expressed or implied, or assumes any legal liability or responsibility for the accuracy, completeness, or usefulness of any information, apparatus, product, or process disclosed, or represents that its use would not infringe privately owned rights. Reference herein to any specific commercial product, process, or service by trade name, trademark, manufacturer, or otherwise does not necessarily constitute or imply its endorsement, recommendation, or favoring by the United States government or Lawrence Livermore National Security, LLC. The views and opinions of authors expressed herein do not necessarily state or reflect those of the United States government or Lawrence Livermore National Security, LLC, and shall not be used for advertising or product endorsement purposes. 
\end{acknowledgments}

\section*{Data Availability Statement}

The data that support the findings of this study are available from the corresponding author upon reasonable request.

\appendix

\section{Detailed Tuning Process}
\label{appendix}

\begin{figure}
\includegraphics[width=\columnwidth]{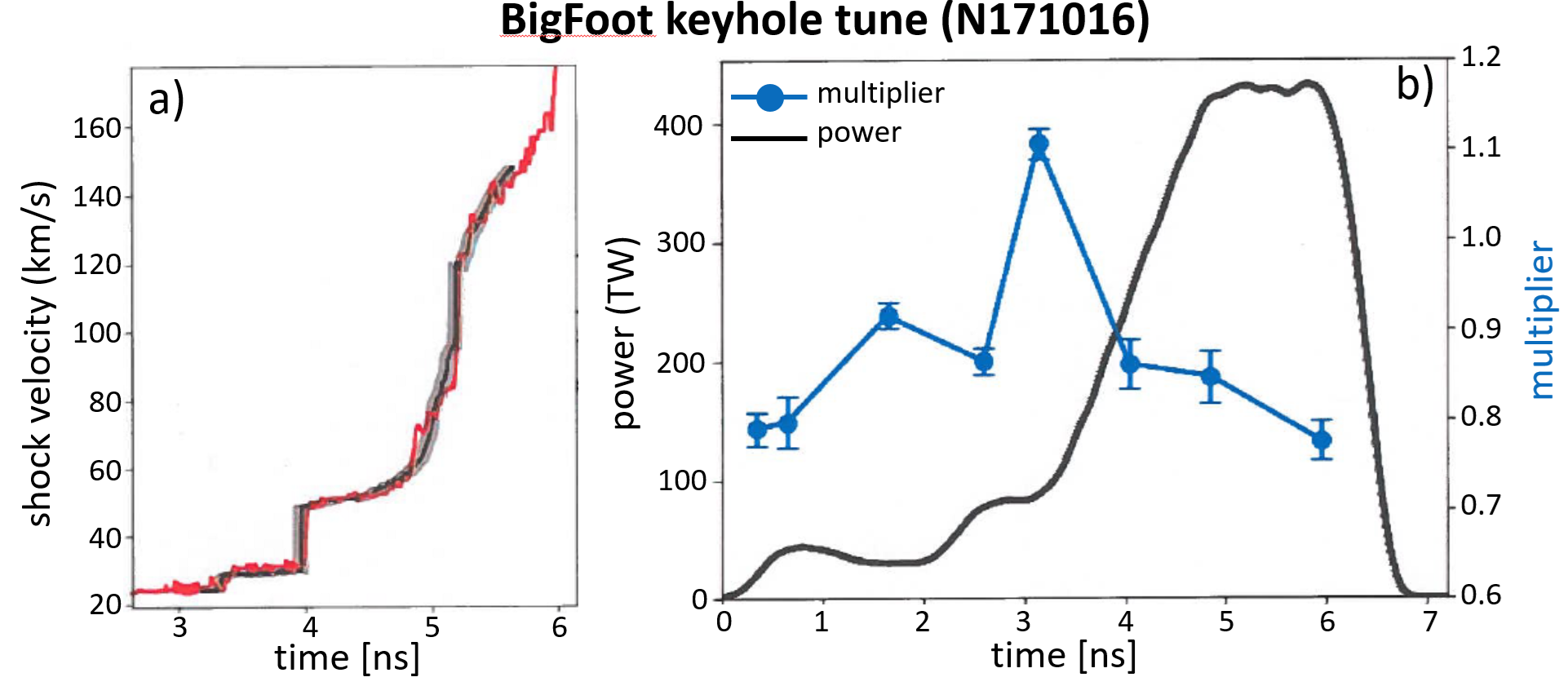}
\caption{\label{fig:shock_multiplier} (a) An example of the leading shock velocity from a simulation (red) fitted to the experimental VISAR data (black) and (b) the final power multipliers (blue) relative to the baseline laser drive used to match shock velocities.}
\end{figure}

The first step of the ANTS tuning process used in this study involves an associated keyhole experiment, for example, keyhole N171016 for DT shot N180128, where the foot of the pulse is stretched out and the peak power is truncated so as to clearly resolve the shock velocities and timings in question using VISAR.\cite{Malone} This sets the laser power multipliers before peak power. The peak power multiplier can be tuned separately since it has little impact on the VISAR data. A small suite of simulations is run where time-power multipliers are varied to sample the total parameter space with a dimensionality of 6 - 8 parameters. 
A depiction of the power multipliers used from the shock tune can be found in Fig. \ref{fig:shock_multiplier}.

\begin{figure}
\includegraphics[width=\columnwidth]{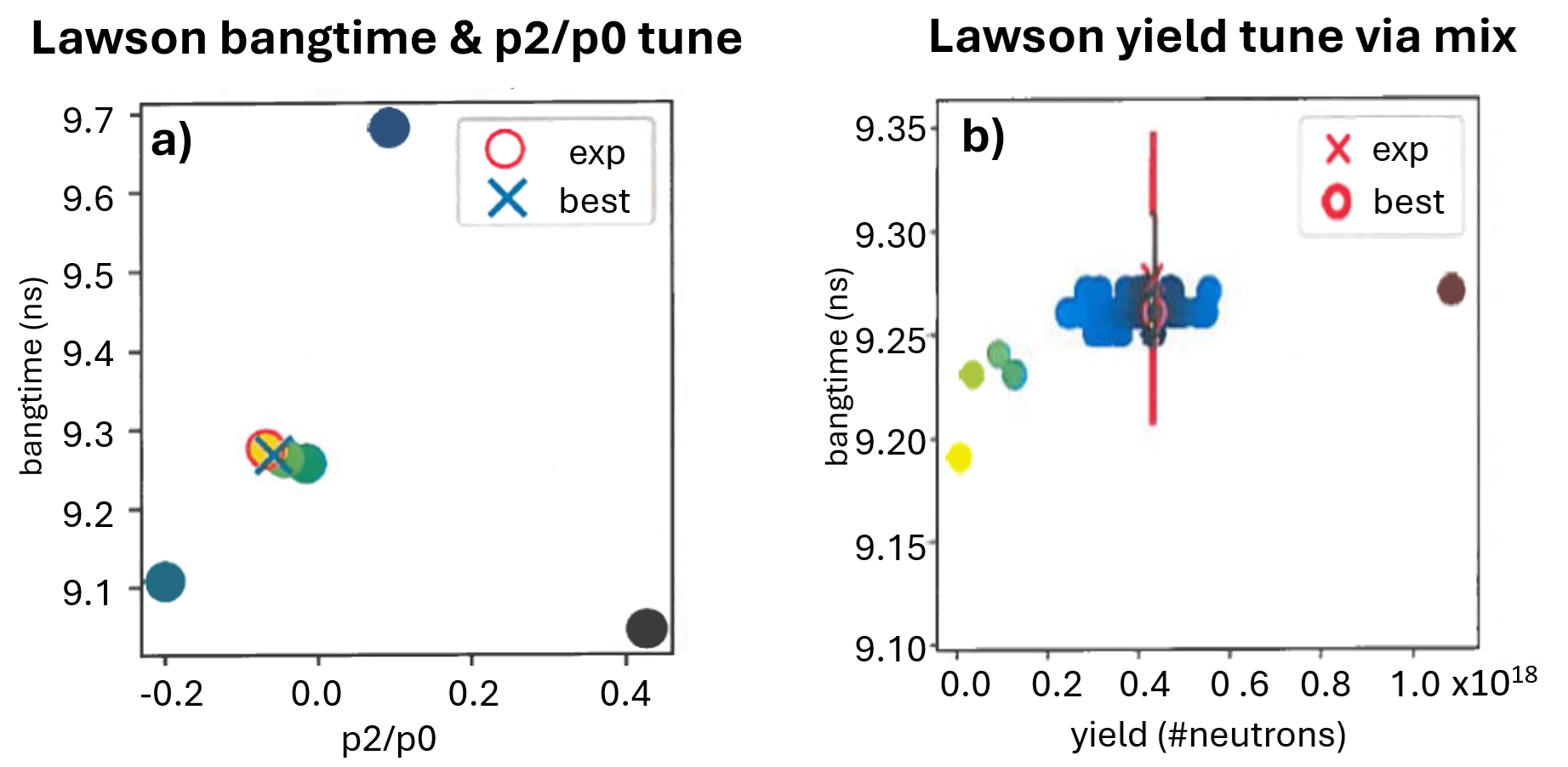}
\caption{\label{fig:btp2} (a) Depiction of the bangtime and $p_2/p_0$ parameter space as fitted by a Bayesian optimization algorithm for shot N210808 and (b) a depiction of the bangtime and yield parameter space as we tuned the mix model to fit bangtime, yield, $p_2/p_0$, and minimize $p_4/p_0$ and $p_6/p_0$.}
\end{figure}

Once the foot power multipliers are fit to the keyhole data, we next tune for bangtime, which is the time of peak neutron production, and $p_2/p_0$ hotspot shape, where $p_2/p_0$ is the ratio of the second-order Legendre polynomial coefficient fit divided by the baseline spherical fit. This entails a new suite of simulations that directly model the DT implosion in question and is much more costly. Tritium-hydrogen-deuterium (THD) shots can be used as well and matched to a "burn-off" simulation, but these typically occur after the experiment and so are not always available. Here, our free parameters are the peak power multiplier, typically less than 1 and often around 0.8, and then a shape factor, where here we primarily used a CBET clamp $\delta n/n$, which starts at a time late in the foot $t_\text{CBET}$. One can also use a cone fraction multiplier, which is the ratio of inner cone energy to total laser energy, i.e., $(E_\text{23+30}/E_\text{total})$. A visualization of the parameter search for the $t_{bt}$ \& $p_2/p_0$ tune can be seen in Fig. \ref{fig:btp2} (a).  Once bangtime and shape are tuned, we typically over predict the yield by around 2x. 

\begin{figure}
\includegraphics[width=\columnwidth]{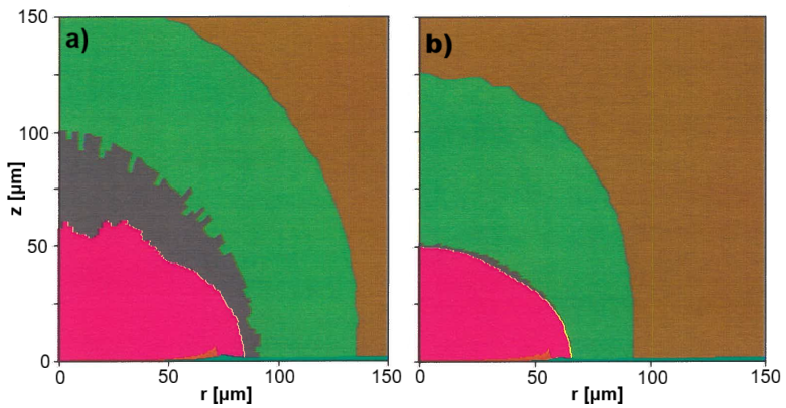}
\caption{\label{fig:mix2} Depictions of (a) N210808 and (b) N221204 with respect to material regions near stagnation. The magenta region defines DT while green and brown represent the ablator. The gray region represents DT-carbon mix from \lasnex{}'s buoyancy-drag model.}
\end{figure}

As the last step we degrade the yield by using a buoyancy-drag mix model that is a function of three parameters: the mix width $h_{mw} [cm]$, which defines an initial mixing amplitude of coupled spikes and bubbles in the mix region, the premix fraction $f_{pmA} [\%]$, which is the atomic fraction of ablator material (carbon or beryllium here) assumed to initially be in the DT vapor region, and the dopant mix fraction multiplier $f_{pmD}$, where the initial dopant (tungsten or molybdenum) fraction in the DT vapor is 
 $f_{pmD}\times f_{pmA}\times f_\text{dopant}$, $f_\text{dopant}$ being listed in Table \ref{tab:tabtarg}. The buoyancy-drag model is governed by an equation of the form:
$$
\frac{dV_i}{dt} = Ag - C \frac{V_i |V_i|}{h_i},
$$
where $V_i = dh_i/dt$, $h_i$ is the time-dependent mix width where $h_i(t=0)=h_{mw}$, $A$ is a coefficient proportional to the Atwood number, i.e., $A\propto (\rho_2-\rho_1)/(\rho_2+\rho_1)$, where $\rho_2$ and $\rho_1$ are the heavy and light fluids respectively, $g$ is the acceleration, and $C$ is the drag coefficient. Mix not only degrades the yield but also alters the shape of the implosion, complicating the final tuning process. In \lasnex{} mix is represented by a distinct "mix region", as visualized in Fig. \ref{fig:mix2}, where in (a) we depict N210808 near stagnation as a case example with greater mix, while in (b) we have a low-mix example of N221204.  Since our problem is under-determined, we typically also try to tune for higher-order shape factors, where $p_4$ is sometimes measured in experiments or if not available to just minimize $p_4$ and $p_6$ in general to improve the shape as much as possible. The final tuned parameters can be found in Table \ref{tab:tune} and then a comparison of tuned simulations to experiments in Table \ref{tab:table1}. A visualization of the parameter search for the $t_{bt}$ \& $Y$ slices of the mix tune can be seen in Fig. \ref{fig:btp2} (b).

Bangtime was relatively simple to match within the error bars of the experiment in addition to the shocks for all three NIF postshot simulations. For one experiment, N221204, the final shock was not measured by the VISAR diagnostic which reduced potential limitations to the fit. To compensate for this the shock timings for N210808 were used to initialize the fits for N221204 given the similarity between the designs. In addition, for N180128 it was not possible to fit bangtime and $p_2/p_0$ with just the peak power multiplier and a CBET clamp $\delta n/n$ and so some minor cone fraction $CF$ was introduced to aid with that. The error tolerances for the simulations relative to the experiment for the target parameters can be found in Table \ref{tab:expbounds}.
\begin{table}
\caption{\label{tab:expbounds} The target bounds within which we tune simulation parameters to experimental data for shots N180128, N210808, and N221204. $p_4/p_0$ data was only available for N210808 and $p_6/p_0$ was not available for any of the experimentally matched shots.}
\begin{ruledtabular}
\begin{tabular}{c|c|c|c|c|c|c}
Exp. parameter & Y/Y$_\text{exp}$ & $t_{bt}$ & \Ti & $p_2/p_0$ & $p_4/p_0$ & $p_6/p_0$  \\
\hline
Tuning threshold & $\pm$25 \% & $\pm$30 ps & $\pm$2 keV &  $\pm$10\% &  n/a &  n/a  \\
\end{tabular}
\end{ruledtabular}
\end{table}

A caveat here is that \Ti is the burn-averaged, birth (thermal) ion temperature in the hot spot, which does not take into account Doppler shifting of the measured neutron spectrum whence the temperature is inferred, which results in a higher measured temperature than the baseline thermal \Ti. In experiments the measured \Ti is higher than that in the simulation, so our predictions should be a bit lower. Lastly, the yield for N221204 was kept high, 44\% more than the experiment, to reduce the time spent tuning as well as to keep pace with recent HYBRID-E shots that had very similar designs but even higher yields (the current record at the time this study was done was N240210 at 5.2 MJ). The case examples of EYC and PSS do not have analogous experimental shots to date thus far, so best practices were assumed with respect to the power multipliers and optimal shape. PSS power multipliers were based on previous studies, e.g., Ref. \onlinecite{maclaren2}, and which inform ongoing and future experimental campaigns. EYC tuning was derived from the simulations used in Ref. \onlinecite{maclaren1}, which was originally tuned using wall opacity multipliers and careful tuning of the shape to optimize shock breakout. This simulation was redone with laser power multipliers, to be consistent with the rest of the simulations in this study, where one power multiplier was used for the foot and one for the peak.

\begin{table*}
\caption{\label{tab:tune} Tuning parameters used to calibrate simulations to experiments in the case of N180128, N210808, and N221204 and then best practices tuning for preshot design of EYC and PSS. Here $t_i$ is laser pulse time [ns], $P_i$ is the associated power multiplier,  $t_\text{CBET}$ is the time the CBET saturation clamp is applied, CBET $\delta n/n$ is the CBET saturation clamp, CF is the laser energy cone fraction, $h_{mw}$ is the initial mix width, $f_{pmA}$ is the initial ablator mix fraction, and $f_{pmD}$ is the dopant layer mix fraction multiplier.}
\begin{ruledtabular}
\begin{tabular}{l|c|c|c|c|c}
Shot & {BigFoot}   & {Lawson} & {Break-even} & Enhanced Yield & Pushered Single\\
   & {(N180128)}  & {(N210808)} & {(N210808)} & Capability (EYC) & Shell (PSS) \\
\hline
Keyhole & N171016    &  N221106    &  N230110    & -   & -  \\
\hline
$t_1$ & 0.4  &  0.5  &  0.5  & 5.0 & 3.9  \\
$t_2$ & 0.7  &  0.75 &  0.75 & - & 5.5  \\ 
$t_3$ & 1.5  &  2.1  &  1.95 & - & 6.15  \\ 
$t_4$ & 2.15  &  3.2 &  3.25 & - & 6.6  \\ 
$t_5$ & 2.65  &  3.7 &  3.85 & - & 7.6  \\
$t_6$ & 3.65  &  4.5 &  4.65 & - & 8.15 \\
$t_\text{peak}$ & 4.65  &  5.25 &  5.4 & 7.1 & 10.0  \\
\hline
$P_1$ & 0.79   &  0.658 &  0.658 & 0.86 & 0.85  \\
$P_2$ & 0.797  &  0.773 &  0.773 & - & 1.00  \\
$P_3$ & 0.916  &  0.836 &  0.836 & - & 0.70  \\
$P_4$ & 0.865  &  0.705 &  0.805 & - & 0.925  \\
$P_5$ & 1.106  &  1.132 &  1.132 & - & 1.20  \\
$P_6$ & 0.862  &  0.699 &  0.699 & - & 0.75  \\
$P_\text{peak}$ & 0.83  &  0.839 &  0.854 & 0.8825 & 0.85  \\
\hline
$t_\text{CBET}$ [ns] & 3.0  & 3.8   & 3.8  &   0  &  0 \\  
CBET $\delta n/n$ & 0.015 & 0.0021 &  0.0028  & 0.0075   & 0.0085  \\   
$CF$ & 0.274 & 0.3  &  0.3    & 0.3   &  0.3 \\  
$h_{mw} [cm]$  & 1.5e-5 & 3.8e-5  &  3.3e-5  & 3.5e-5 &  2.5e-5 \\
$f_{pmA} [\%]$ & 0.0075 & 0.0095  &  0.06    & 0.01   &  0.02 \\  
$f_{pmD}$   & 0.015  &  0.6    & 0.018    & 0.2    &  0.02 \\   
\end{tabular}
\end{ruledtabular}
\end{table*}

\nocite{*}
\bibliography{aipsamp}

\end{document}